\documentclass[aps,pre,twocolumn,groupedaddress]{revtex4-2}
\usepackage{bm}
\usepackage{dcolumn}
\usepackage{braket}
\usepackage{graphicx}
\usepackage{amssymb}
\usepackage{amsmath}

\begin{document}
\title{Efficient quantum and simulated annealing of Potts models using a half-hot constraint}

\author{Shuntaro Okada$^{1,2}$}
\author{Masayuki Ohzeki$^{1,3,4,5}$}
\author{Kazuyuki Tanaka$^{1}$}
\affiliation{$^1$Graduate School of Information Sciences, Tohoku University, Sendai 980-8579, Japan}
\affiliation{$^2$Electronics R \& I Division, DENSO CORPORATION, Tokyo 103-6015, Japan}
\affiliation{$^3$Institute of Innovative Research, Tokyo Institute of Technology, Yokohama 226-8503, Japan}
\affiliation{$^4$Jij Inc., Tokyo 113-0033, Japan}
\affiliation{$^5$Sigma-i Co. Ltd., Tokyo 108-0075, Japan}

\date{\today}

\begin{abstract}
The Potts model is a generalization of the Ising model with $Q>2$ components.
In the fully connected ferromagnetic Potts model, a first-order phase transition is induced by varying thermal fluctuations.
Therefore, the computational time required to obtain the ground states by simulated annealing exponentially increases with the system size.
This study analytically confirms that the transverse magnetic-field quantum annealing induces a first-order phase transition.
This result implies that quantum annealing does not exponentially accelerate the ground-state search of the ferromagnetic Potts model.
To avoid the first-order phase transition, we propose an iterative optimization method using a half-hot constraint that is applicable to both quantum and simulated annealing.
In the limit of $Q \to \infty$, a saddle point equation under the half-hot constraint is identical to the equation describing the behavior of the fully connected ferromagnetic Ising model, thus confirming a second-order phase transition.
Furthermore, we verify the same relation between the fully connected Potts glass model and the Sherrington--Kirkpatrick model under assumptions of static approximation and replica symmetric solution.
The proposed method is expected to obtain low-energy states of the Potts models with high efficiency using Ising-type computers such as the D-Wave quantum annealer and the Fujitsu Digital Annealer.
\end{abstract}
\maketitle

\section{Introduction}
Combinatorial optimization problems, which minimize the cost function with discrete variables, have significant real-world applications.
Generally, the cost function of a combinatorial optimization problem can be mapped to the Hamiltonian of a classical Ising model \cite{Ising_mapping}.
Simulated annealing (SA) \cite{SA_original} is a heuristic algorithm that searches the ground state of a Hamiltonian, exploiting thermal fluctuations to escape the local minima.
In contrast to SA, quantum annealing (QA) \cite{QA_original}, which is strongly related to adiabatic quantum computation \cite{AQC_original}, escapes the local minima through the tunneling effects induced by quantum fluctuations, that are usually imposed by a transverse magnetic field.
At the beginning of QA, a strong transverse magnetic field is applied, and the system is set in the trivial ground state with all spins aligned along the transverse magnetic field.
Subsequently, the transverse magnetic field is reduced to zero, and the system evolves according to the Schr\"{o}dinger equation.
If the system changes sufficiently slowly, it remains close to the instantaneous ground state of the time-dependent Hamiltonian.
According to the adiabatic theorem \cite{adiabatic_theorem}, the computational time of QA is proportional to the inverse square of the minimum energy gap between the instantaneous gound state and the first excited state.
Whether quantum effects accelerate the computation of the ground-state search is one of the primary research topics that has been discussed in several studies \cite{QA_SA_compare1, QA_SA_compare2, QA_SA_compare3, QA_SA_compare4, QA_SA_compare5, QA_SA_compare6}.
Moreover, further improvements to QA have also been theoretically discussed.
One of the promising directions of improvements is implementing the XX interaction and introducing a non-stoquastic Hamiltonian.
With the exception of certain cases \cite{non-stoquastic_QMC}, the XX interaction and other non-trivial quantum fluctuations hinder efficient classical computations because of a sign problem.
However, the non-trivial quantum fluctuations can accelerate the computation of several specific problems \cite{XX_p-spin, XX_Hopfield, XX_p-spin_gap, XX_QMC, XX_one-dimension}.
Hence, the introduction of non-trivial quantum fluctuations might be essential for achieving quantum supremacy and for boosting the power of QA.

Recently, D-Wave Systems Inc. developed commercial QA machines based on superconducting flux qubits \cite{D-Wave_machine}.
The performances of QA and SA has been compared in experimental studies on the D-Wave quantum annealer \cite{D-Wave_compare1, D-Wave_compare2, D-Wave_compare3},
and the applicability of the quantum annealer to practical problems has been demonstrated \cite{D-Wave_application1, D-Wave_application2, D-Wave_application3, D-Wave_application4, D-Wave_application5, D-Wave_application6, D-Wave_application7, D-Wave_application8, D-Wave_application9, D-Wave_application10, D-Wave_application11, D-Wave_application12, D-Wave_application13, D-Wave_application14, D-Wave_application15, D-Wave_application16}.
Although many cost functions in practical problems are naturally mapped to the Hamiltonian of a Potts model \cite{Potts} rather than that of an Ising model,
the D-Wave quantum annealer requires that the cost function is represented in the same form as the Ising model.
Generally, one-hot encoding is employed to represent the ground-state search of the Potts model as that of the Ising model, and it is widely applicable to the optimization performed by the Ising-type computers such as D-Wave quantum annealer and Fujitsu Digital Annealer \cite{digital_annealer}.
Nevertheless, to the best of our knowledge, the performance of QA and SA with one-hot encoding has not been adequately investigated.

In this study, we focuse on the ground-state search of the Potts models with one-hot encoding.
We analytically investigated the phase-transition order during QA in the fully connected ferromagnetic (FM) Potts model and confirmed the occurrence of the first-order phase transition.
In a system with first-order phase transitions, the minimum energy gap typically decreases exponentially with the system size \cite{gap_first-order1, gap_first-order2, gap_first-order3}, indicating that QA cannot efficiently idnetify a ground state,
while the minimum energy gap decreases polynomially in a system with second-order phase transitions.
Therefore, it is conjectured that the computational time of QA in the ground-state search of the FM Potts model exponentially increases with the system size, as with SA.
Subsequently, to avoid the first-order phase transition, we propose an iterative optimization method under a half-hot constraint that is applicable to both QA and SA.
Under the half-hot constraint and in the limit of $Q \to \infty$, the saddle point equation of the FM Potts model is identical to that of the FM Ising model, indicating a second-order phase transition.
We further confirm the same relation between the fully connected Potts glass (PG) model \cite{Potts_glass} and the Sherrington-Kirkpatrick (SK) \cite{SK_model1, SK_model2} model by assuming the static approximation and the replica symmetric solution.
According to these results, by introducing the half-hot constraint, the difficulty of obtaining the ground states of a Potts model might be generally reducible to that of obtaining ground states of the corresponding Ising model.

The remainder of this study is organized as follows.
In Section II, we briefly explain one-hot encoding for the Hamiltonian of Potts models.
In Section III, we present the verification of the first-order phase transition during QA of the FM Potts model and propose an iterative optimization under the half-hot constraint to avoid the first-order phase transition.
In Section IV, we investigate the iterative optimization of the fully connected PG model under the half-hot constraint.
Finally, in Section V, we present the discussion and conclusion of this study.

\section{One-hot encoding for the Hamiltonian of Potts models}
\begin{figure}
	\includegraphics[width=8.6cm]{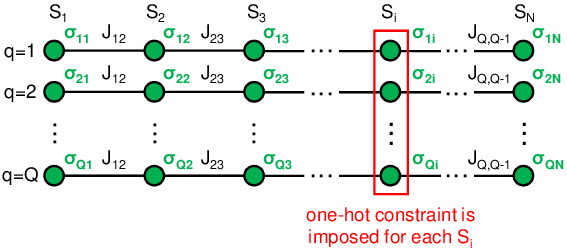}
	\caption{Encoded Ising model for a one-dimensional Potts model.
	Vertices and edges represent Ising spins and interactions between them, respectively.
	Although the penalty term generates fully connected interactions between $\sigma_{qi}$ and $\sigma_{q'i}$, they are not shown for simplicity.
	$Q$ Ising spins $\{ \sigma_{qi} \}_{q=1, 2, ..., Q}$ are assigned to each $S_{i}$, and only one spin is allowed to be $-1$ among the $Q$ Ising spins for each $S_{i}$.}
	\label{fig:Ising_example}
\end{figure}

In this section, we briefly explain one-hot encoding for the Hamiltonian of Potts models.
The Hamiltonian investigated in this study is as follows:
\begin{equation}
\mathcal{H}_{\mathrm{potts}} = - \frac{4}{N} \sum_{i<j}^{N} J_{ij} \delta \left( S_{i}, S_{j} \right),  \label{eq:Potts_original}
\end{equation}
where $S_{i} \in \left( 1, 2, ..., Q \right)$ is a Potts spin with $Q$ components, $N$ represents the number of Potts spins, $J_{ij}$ is an interaction between two Potts spins, and $\delta$ is the Kronecker delta function.
The equivalent binary optimization problem with one-hot encoding is given by the following equation:
\begin{equation}
\underset{\bm{x}}{\mathrm{argmin}} \left[ - \frac{4}{N} \sum_{i<j}^{N} J_{ij} \sum_{q=1}^{Q} x_{qi} x_{qj} \right] \ \  \mathrm{s.t.} \ \  \sum_{q=1}^{Q} x_{qi} = 1,  \label{eq:OH_constrained}
\end{equation}
where $x_{qi} \in (0,1)$ is the binary variable assigned to the component $q$ of $S_{i}$, $x_{qi}=1$ indicates that the component $q$ is selected for $S_{i}$, and $J_{ij}$ contributes to the energy only when $x_{qi} = x_{qj} = 1$.
The constraint, which we call ``one-hot constraint'', restricts feasible solutions to configurations where exactly one component is selected for each $S_{i}$.
Thus, the Potts spin $S_{i}$ is given by the following equation:
\begin{equation}
S_{i} = \sum_{q=1}^{Q} q x_{qi}.
\end{equation}
Using the transformation:
\begin{equation}
x_{qi} = \frac{ 1 - \sigma_{qi} }{2},  \label{eq:binary2Ising}
\end{equation}
the abovementioned optimization problem [Eq. (\ref{eq:OH_constrained})] can be expressed with respect to the Ising spin $\sigma_{qi} \in (+1, -1)$ as follows:
\begin{eqnarray}
\underset{\bm{\sigma}}{\mathrm{argmin}} \left[ - \frac{1}{N} \sum_{i<j} J_{ij} \sum_{q=1}^{Q} \sigma_{qi} \sigma_{qj} \right] &\ &  \nonumber  \\
\mathrm{s.t.} \ \  \sum_{q=1}^{Q} \sigma_{qi} &=& Q-2,
\end{eqnarray}
where $\sigma_{qi} = -1$ indicates that the component $q$ is selected for $S_{i}$.
Here, we neglect the first-order terms proportional to $\sum_{q=1}^{Q} \sigma_{qi}$ because they are constrained to a constant ($Q-2$).
The unconstrained cost function, which is required for optimization by the D-Wave quantum annealer, is obtained by introducing the following penalty term:
\begin{eqnarray}
\mathcal{H}_{0} = &-& \frac{1}{N} \sum_{i<j}^{N} J_{ij} \sum_{q=1}^{Q} \sigma_{qi} \sigma_{qj}  \nonumber  \\
&\ & \ \ \ \ \  + \frac{\lambda}{2Q} \sum_{i=1}^{N} \left[ \sum_{q=1}^{Q} \sigma_{qi} - (Q-2) \right]^{2},  \label{eq:one-hot_Ising}
\end{eqnarray}
where the second term is the penalty term, which is minimized only when the one-hot constraint is satisfied, and the parameter $\lambda$ controls the strength of the penalty term.
By setting the parameter $\lambda$ to a sufficiently large value, the ground states of the Ising model [Eq. (\ref{eq:one-hot_Ising})] correspond to those of the original Potts model.
The one-dimensional Potts model is encoded to the Ising model shown in Fig. \ref{fig:Ising_example}.

\begin{figure*}
	\includegraphics[width=2.0\columnwidth]{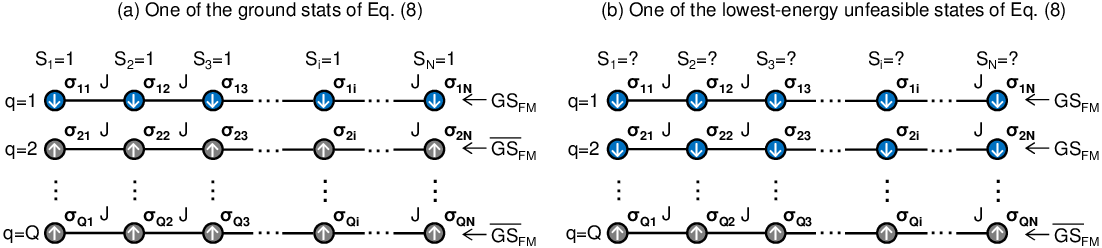}
	\caption{Vertices and edges represent Ising spins and interactions between them, respectively.
	For simplicity, only nearest-neighbor interactions caused by the first term of Eq. (\ref{eq:ferro_one-hot_rep}) are shown.
	Upward and downward arrows indicate $\sigma^{(z)}_{qi} = +1$ and $\sigma^{(z)}_{qi} = -1$, respectively.
	$\mathrm{GS}_{\mathrm{FM}}$ and $\overline{\mathrm{GS}}_{\mathrm{FM}}$ denote the ground states of the FM Ising model, and the spins in $\overline{\mathrm{GS}}_{\mathrm{FM}}$ are all reversed from $\mathrm{GS}_{\mathrm{FM}}$.
	(a) One of the ground states of Eq. (\ref{eq:ferro_one-hot_rep}), where the component $q=1$ is commonly selected, namely, $S_{i}=1$ for all $i$.
	(b) One of the lowest-energy unfeasible states of Eq. (\ref{eq:ferro_one-hot_rep}), where the components $q=1$ and $q=2$ are commonly ``selected''.
	In this case, $S_{i}$ cannot be determined.}
	\label{fig:ferro_gound_unfeasible}
\end{figure*}
\section{Annealing of the ferromagnetic Potts model}
This section analytically investigates the order of the phase transition during QA in the fully connected FM Potts model and proposes an iterative optimization under the half-hot constraint.
The proposed method is applicable to both QA and SA.
The Hamiltonian of QA is given as follows:
\begin{equation}
\hat{\mathcal{H}} = \hat{\mathcal{H}}_{0} + \hat{\mathcal{H}}_{\mathrm{q}},  \label{eq:ferro_H0_Hq}
\end{equation}
\begin{eqnarray}
\hat{\mathcal{H}}_{0} &=& - \frac{J}{2N} \sum_{q=1}^{Q} \left( \sum_{i=1}^{N} \hat{\sigma}^{(z)}_{qi} \right)^{2}  \nonumber  \\
&+& \frac{\lambda}{2Q} \sum_{i=1}^{N} \left[ \left( \sum_{q=1}^{Q} \hat{\sigma}^{(z)}_{qi} \right)^{2} - 2(Q-2) \sum_{q=1}^{Q} \hat{\sigma}^{(z)}_{qi} \right],  \label{eq:ferro_one-hot_rep}
\end{eqnarray}
\begin{equation}
\hat{\mathcal{H}}_{\mathrm{q}} = - \Gamma \sum_{i=1}^{N} \sum_{q=1}^{Q} \hat{\sigma}^{(x)}_{qi},  \label{eq:ferro_one-hot_fluctuation}
\end{equation}
where $J_{ij}$ is set to $J > 0$, $\Gamma$ controls the strength of the transverse magnetic field, and $\sigma^{(z)}_{qi}$ and $\sigma^{(x)}_{qi}$ are the Pauli $z$ and $x$ operators, respectively.
The second term in Eq. (\ref{eq:one-hot_Ising}) is expanded, and the constant term is neglected in Eq. (\ref{eq:ferro_one-hot_rep}).
The penalty term consists of a fully connected anti-FM interaction and longitudinal magnetic field.
One of the ground states and lowest-energy unfeasible states of Eq. (\ref{eq:ferro_one-hot_rep}) are shown in Fig. \ref{fig:ferro_gound_unfeasible}.
The first term in Eq. (\ref{eq:ferro_one-hot_rep}) represents the sum of the Hamiltonians of the $Q$ independent FM Ising models, and the spin configurations shown in Fig. \ref{fig:ferro_gound_unfeasible} comprise the ground states of the FM Ising model, $\mathrm{GS}_{\mathrm{FM}}$, and $\overline{\mathrm{GS}}_{\mathrm{FM}}$.
The first term in Eq. (\ref{eq:ferro_one-hot_rep}) is obviously minimized in both states shown in Fig. \ref{fig:ferro_gound_unfeasible}, and the difference of the energy is caused by the penalty term, which is equal to $2N \lambda /Q$.
Therefore, $\lambda >0$ is sufficient to correctly encode the ground states of the FM Potts model.
In other words, unfeasible states cannot be the ground states of Eq. (\ref{eq:ferro_one-hot_rep}) when $\lambda>0$ .

\subsection{QA under the one-hot constraint}
In this subsection, we confirm that a first-order phase transition occurs during QA when $Q > 2$.
Using the Suzuki--Trotter formula \cite{ST_formula} and the static approximation, which assumes constancy along the Trotter slice, we obtain free energy in the limit of $N \to \infty$ and $\beta \to \infty$ (see appendix A for detailed calculations) as follows:
\begin{equation}
f ( \{ m_{q} \} ) = \frac{J}{2} \sum_{q=1}^{Q} m_{q}^{2} + \varepsilon^{(\mathrm{eff})}_{\mathrm{min}} ( \{ m_{q} \} ),  \label{eq:ferro_OH_free1}
\end{equation}
where $m_{q}$ is the FM order parameter for $\{ \sigma^{(z)}_{qi} | i = 1, 2, ..., N \}$, $\varepsilon^{(\mathrm{eff})}_{\mathrm{min}} ( \{ m_{q} \} )$ is the lowest eigenvalue of $\hat{\mathcal{H}}^{(\mathrm{eff})}$, and $\hat{\mathcal{H}}^{(\mathrm{eff})}$ is given by the following equation:
\begin{eqnarray}
\hat{\mathcal{H}}^{(\mathrm{eff})} && ( \{ m_{q} \} ) = \frac{\lambda}{2Q} \left( \sum_{q=1}^{Q} \hat{\sigma}^{(z)}_{q} \right)^{2}  \nonumber  \\
	&&- \sum_{q=1}^{Q} \left( J m_{q} + \frac{Q-2}{Q} \lambda \right) \hat{\sigma}^{(z)}_{q} - \Gamma \sum_{q=1}^{Q} \hat{\sigma}^{(x)}_{q}.  \label{eq:ferro_OH_free2}
\end{eqnarray}
We can numerically calculate $\varepsilon^{(\mathrm{eff})}_{\mathrm{min}} ( \{ m_{q} \} )$ for small $Q$, and the order parameters $\{ m_{q} \}$ are determined as the minimizer of free energy.
Because the one-hot constraint is imposed in $\hat{\mathcal{H}}_{0}$, it is reasonable to assume that one order parameter is equal to $m^{(-)}$, and the rest of $Q-1$ order parameters are equal to $m^{(+)}$ at the global minimum of the free energy.
The trivial ground state at $\Gamma = 0$ is $m^{(-)} = -1$ and $m^{(+)} = +1$, as shown in Fig. \ref{fig:ferro_gound_unfeasible}(a).

\begin{figure}
	\includegraphics[width=8.6cm]{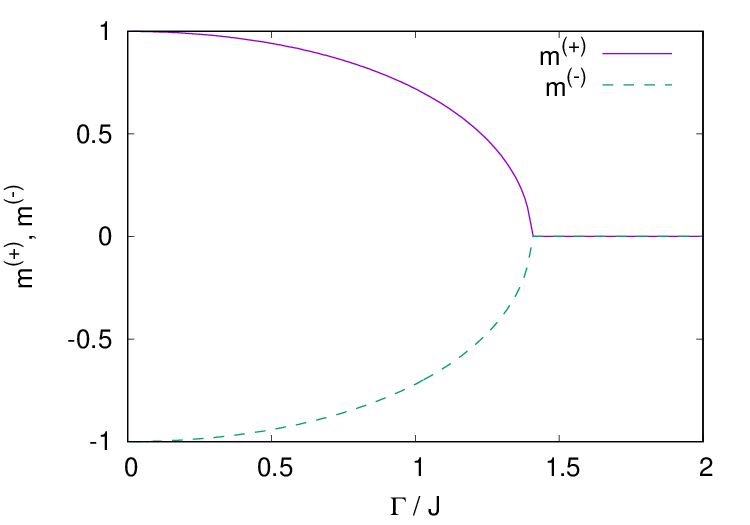}
	\caption{Order parameters $m^{(\pm)}$ for the fully connected FM Potts model with $Q=2$ components.
	As $\Gamma$ decreases, the order parameters $m^{(\pm)}$ continuously change from $m^{(\pm)} = 0$ to $m^{(\pm)} \gtrless 0$.}
	\label{fig:ferro_Q=2_lambda=1}
\end{figure}

\begin{figure}
	\includegraphics[width=8.6cm]{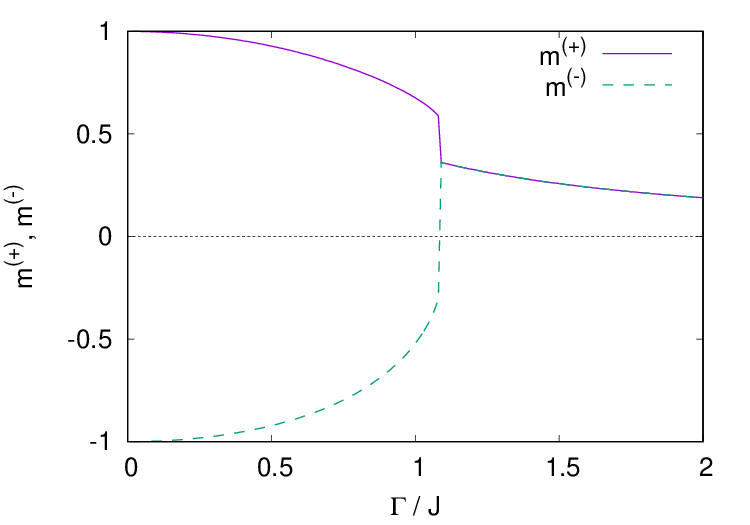}
	\caption{Order parameters $m^{(\pm)}$ for the fully connected FM Potts model with $Q=3$ components.
	The order parameters are biased to $m^{(\pm)}>0$ at a large $\Gamma$. As $\Gamma$ decreases, the order parameters $m^{(\pm)}$ discontinuously change from $m^{(\pm)}>0$ to $m^{(\pm)} \gtrless 0$.}
	\label{fig:ferro_Q=3_lambda=1}
\end{figure}

\begin{figure}
	\includegraphics[width=8.6cm]{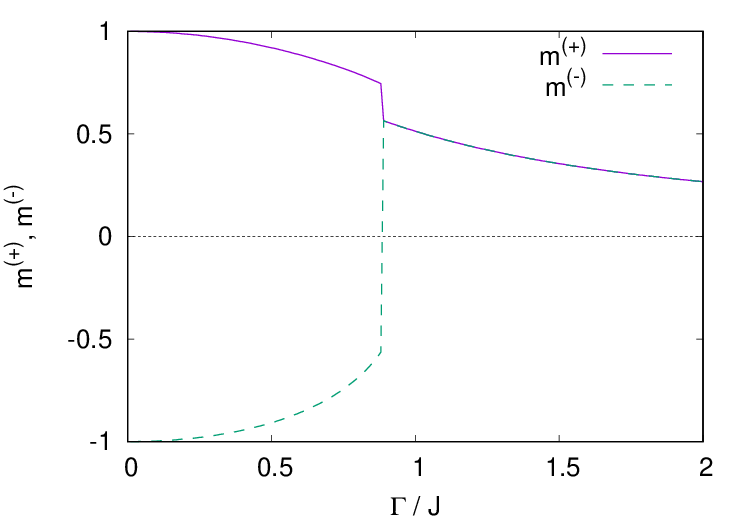}
	\caption{Order parameters $m^{(\pm)}$ for the fully connected FM Potts model with $Q=4$ components.
	The order parameters are biased to $m^{(\pm)}>0$ at a large $\Gamma$. As $\Gamma$ decreases, the order parameters $m^{(\pm)}$ discontinuously change from $m^{(\pm)}>0$ to $m^{(\pm)} \gtrless 0$.}
	\label{fig:ferro_Q=4_lambda=1}
\end{figure}

The order parameters $m^{(\pm)}$ that minimize the free energy as functions of $\Gamma$ for $Q=$ 2, 3, and 4 are plotted in Figs. \ref{fig:ferro_Q=2_lambda=1}, \ref{fig:ferro_Q=3_lambda=1}, and \ref{fig:ferro_Q=4_lambda=1}, respectively.
The parameter $\lambda / J$ is set to $1$.
For a small $\Gamma$, we find the order parameters to be $m^{(+)} > 0$ and $m^{(-)} < 0$ for all the three cases.
On the other hand, for a large $\Gamma$, they are $m^{(\pm)} = 0$ for $Q=2$ and $m^{(\pm)} > 0$ for $Q > 2$.
Hence, we can conclude that in between, they must make an abrupt discontinuous jump for $Q > 2$, which is the first-order phase transition, whereas a continuous change occurs for $Q=2$.
These results imply that the computational time of QA exponentially increases with the system size $N$ for the fully connected FM Potts model, as with SA.

The first-order phase transition is caused by the positively biased $m^{(\pm)}$ at a large $\Gamma$, and $m^{(\pm)}$ are biased by the longitudinal magnetic field of the penalty term, whose strength is  $\lambda (Q-2)/Q$.
Whereas, the longitudinal magnetic field is equal to zero for $Q=2$, where the second-order phase transition occurs.

\subsection{Iterative optimization under the half-hot constraint}
To avoid the first-order phase transition, we propose an iterative optimization under the half-hot constraint whose penalty term does not contain the longitudinal magnetic field.

First, we introduce the half-hot constraint as follows:
\begin{equation}
\sum_{q=1}^{Q} \sigma_{qi} = 0.
\end{equation}
The penalty term of the half-hot constriant is given by
\begin{equation}
\frac{\lambda}{2Q} \left( \sum_{q=1}^{Q} \sigma_{qi} \right)^{2},
\end{equation}
which comprises only the anti-FM interaction and does not contain the longitudinal magnetic field.
Therefore, the first-order phase transition is expected to be avoided under the half-hot constraint.
Under the one-hot constraint, feasible solutions are restricted to spin configurations where only one Ising spin is equal to $-1$ among the $Q$ spins assigned to each $S_{i}$, meaning that exactly one component is selected in the feasible solution.
Whereas, under the half-hot constraint, $Q/2$ spins are equal to $-1$ among the $Q$ spins in the feasible solution, meaning that $Q/2$ components are extracted.
We regard the extracted $Q/2$ components as candidates of the optimal solutions, and iterate the optimization under the half-hot constraint among the extracted components until one component is selected for each $S_{i}$.

In this scenario, the following optimization problem is solved in the first iteration:
\begin{equation}
\underset{\bm{\sigma}^{(1)}}{\mathrm{argmin}} \left[ - \frac{J}{2N} \sum_{q=1}^{Q} \left( \sum_{i=1}^{N} \sigma^{(1)}_{qi} \right)^{2} \right] \  \mathrm{s.t.} \  \sum_{q=1}^{Q} \sigma^{(1)}_{qi} = 0.  \label{eq:ferro_first_iteration}
\end{equation}
Here, $\sigma^{(1)}_{qi}$ represents the spin variable in the first iteration, and the half-hot constraint is imposed, which is equivalent to $\sum_{q} x^{(1)}_{qi} = Q/2$.
Assume that we obtain $\{ \sigma^{(1)}_{qi} = -1 | q \in ( \mu^{(1)}_{1i}, \mu^{(1)}_{2i}, ..., \mu^{(1)}_{Q/2,i} ) \}$ and $\{ \sigma^{(1)}_{qi} = +1 | q \in ( \nu^{(1)}_{1i}, \nu^{(1)}_{2i}, ..., \nu^{(1)}_{Q/2,i} ) \}$ as the solution,
where $\bm{\mu}^{(1)}_{i}$ and $\bm{\nu}^{(1)}_{i}$ represent the extracted and not extracted components, respectively, for each $S_{i}$ in the first iteration.
Then, in the second iteration, $\{ \sigma^{(1)}_{qi} | q \in \bm{\nu}^{(1)}_{i} \}$ are fixed to $+1$, and an optimal solution under the half-hot constraint is searched among $\{ \sigma^{(1)}_{qi} | q \in \bm{\mu}^{(1)}_{i} \}$.
The optimization problem in the second iteration is given by the following equation:
\begin{eqnarray}
&&\underset{\bm{\sigma}^{(2)}}{\mathrm{argmin}} \left[ - \frac{J}{N} \sum_{i<j} \sum_{q, q'=1}^{Q/2} \delta ( \mu^{(1)}_{qi}, \mu^{(1)}_{q'j} ) \sigma^{(2)}_{qi} \sigma^{(2)}_{q'j} \right.  \nonumber  \\
&&\left. - \frac{J}{N} \sum_{i \neq j} \sum_{q, q'=1}^{Q/2} \delta ( \mu^{(1)}_{qi}, \nu^{(1)}_{q'j} ) \sigma^{(2)}_{qi} \right] \mathrm{s.t.} \sum_{q=1}^{Q/2} \sigma^{(2)}_{qi} = 0,  \label{eq:ferro_second_iteration}
\end{eqnarray}
where $\sigma^{(2)}_{qi} \equiv \sigma^{(1)}_{\mu^{(1)}_{qi},i}$ is the spin variable in the second iteration, and $\delta$ is the Kronecker delta function  (see Appendix B for a detailed derivation of the cost function).
The first term of the cost function represents interactions between the spins assigned to $\bm{\mu}^{(1)}_{i}$ and $\bm{\mu}^{(1)}_{j}$, and the second term represents the longitudinal magnetic field caused by the spins assigned to $\bm{\nu}^{(1)}_{j}$.
By iterating the optimization under the half-hot constraint, we finally obtain one component for each $S_{i}$.

\begin{figure}
	\includegraphics[width=8.6cm]{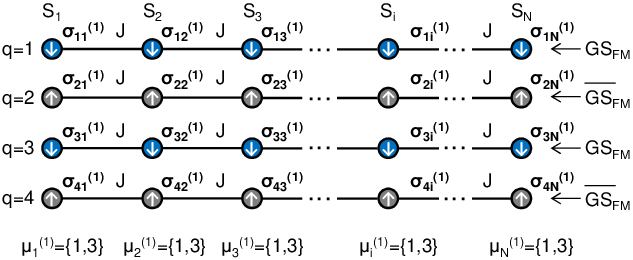}
	\caption{One of the ground states of Eq. (\ref{eq:ferro_first_iteration}) for $Q=4$.
	Vertices and edges represent Ising spins and interactions between them, respectively.
	For simplicity, only the nearest-neighbor interactions are shown.
	$\mathrm{{GS}_{FM}}$ and $\overline{\mathrm{GS}}_{\mathrm{FM}}$ represent the ground states of the FM Ising model, and Ising spins in  $\overline{\mathrm{GS}}_{\mathrm{FM}}$ are all reversed from $\mathrm{GS}_{\mathrm{FM}}$.}
	\label{fig:ferro_HH_ground}
\end{figure}

If each optimization is successfully solved, it is possible to retrieve the ground states of the FM Potts model.
One of the ground states in the first iteration for $Q=4$ is shown in Fig. \ref{fig:ferro_HH_ground}, where $\bm{\mu}^{(1)}_{i} = \{ 1, 3 \}$ and $\bm{\nu}^{(1)}_{i} = \{ 2, 4 \}$ for all $i$.
The ground state comprises those of the FM Ising model, $\mathrm{{GS}_{FM}}$, and $\overline{\mathrm{GS}}_{\mathrm{FM}}$.
The cost function in the first iteration is the sum of the Hamiltonians of the $Q$ independent FM Ising models, which are minimized by the spin configuration shown in Fig. \ref{fig:ferro_HH_ground}.
Furthermore, to satisfy the half-hot constraint, the spin configurations assigned to $q = 2$ and $4$ are all reversed from those assigned to $q=1$ and $3$.
By substituting $\bm{\mu}^{(1)}_{i} = \{ 1, 3 \}$ and $\bm{\nu}^{(1)}_{i} = \{ 2, 4 \}$ into Eq. (\ref{eq:ferro_second_iteration}), we obtain the following optimization problem in the second iteration:
\begin{equation}
\underset{\bm{\sigma}^{(2)}}{\mathrm{argmin}} \left[ - \frac{J}{2N} \sum_{q=1}^{Q/2} \left( \sum_{i=1}^{N} \sigma^{(2)}_{qi} \right)^{2} \right] \ \  \mathrm{s.t.} \ \  \sum_{q=1}^{Q/2} \sigma^{(2)}_{qi} = 0.
\end{equation}
This is equivalent to the optimization of the FM Potts model with $Q/2$ components under the half-hot constraint.
Thus, we can obtain one of the ground states of the FM Potts model by iterating the optimization under the half-hot constraint.

\begin{figure}
	\includegraphics[width=8.6cm]{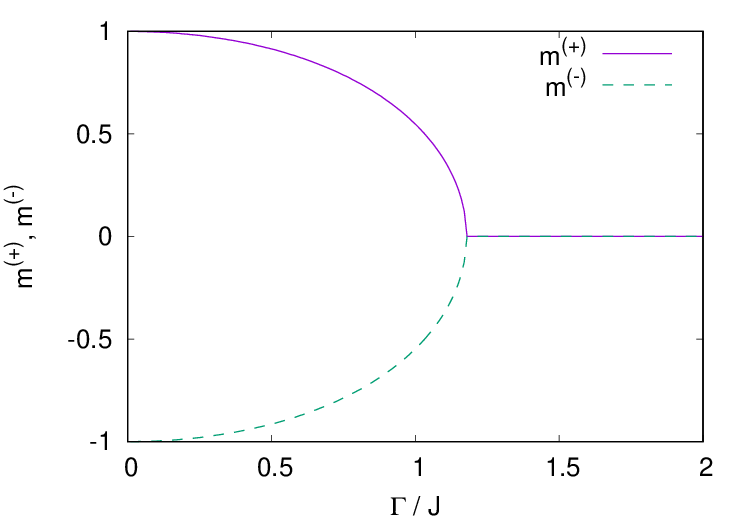}
	\caption{Order parameters $m^{(\pm)}$ for the fully connected FM Potts model with $Q=4$ components under the half-hot constraint.
	Order parameters are not biased at a large $\Gamma$, and the first-order phase transition is avoided.}
	\label{fig:ferro_Q=4_lambda=1_half}
\end{figure}

Next, we show that the first-order phase transition can be avoided under the half-hot constraint for $Q=4$.
The unconstrained cost function is given by the following equation:
\begin{equation}
\hat{\mathcal{H}}_{0} = - \frac{J}{2N} \sum_{q=1}^{Q} \left( \sum_{i=1}^{N} \hat{\sigma}^{(z)}_{qi} \right)^{2} + \frac{\lambda}{2Q} \sum_{i=1}^{N} \left( \sum_{q=1}^{Q} \hat{\sigma}^{(z)}_{qi} \right)^{2}.  \label{eq:ferro_H0_half}
\end{equation}
Note that the penalty term does not contain the longitudinal magnetic field.
Based on the method used to derive Eq. (\ref{eq:ferro_OH_free1}) and Eq. (\ref{eq:ferro_OH_free2}), we obtain the free energy in the limit of $N \to \infty$ and $\beta \to \infty$ as follows:
\begin{equation}
f = \frac{J}{2} \sum_{q=1}^{Q} m_{q}^{2} + \varepsilon^{(\mathrm{eff})}_{\mathrm{min}} ( \{ m_{q} \} ),  \\
\end{equation}
where $\varepsilon^{(\mathrm{eff})}_{\mathrm{min}} ( \{ m_{q} \} )$ is the lowest eigenvalue of $\hat{\mathcal{H}}^{(\mathrm{eff})}$, and $\hat{\mathcal{H}}^{(\mathrm{eff})}$ is given by the following equation:
\begin{eqnarray}
\hat{\mathcal{H}}^{(\mathrm{eff})} ( \{ m_{q} \} ) &=& \frac{\lambda}{2Q} \left( \sum_{q=1}^{Q} \hat{\sigma}^{(z)}_{q} \right)^{2}  \nonumber  \\
&-& J \sum_{q=1}^{Q} m_{q} \hat{\sigma}^{(z)}_{q} - \Gamma \sum_{q=1}^{Q} \hat{\sigma}^{(x)}_{q}.
\end{eqnarray}
As is the case under the one-hot constraint, we assume that $Q/2$ order parameters are equal to $m^{(-)}$, and the rest of $Q/2$ order parameters are equal to $m^{(+)}$ at the global minimum of free energy.
The trivial ground state at $\Gamma = 0$ is $m^{(-)} = -1$ and $m^{(+)} = +1$ as shown in Fig. \ref{fig:ferro_HH_ground}.
The order parameters $m^{(\pm)}$, which minimize the free energy, are plotted as functions of $\Gamma$ in Fig. \ref{fig:ferro_Q=4_lambda=1_half}.
They are not positively biased at a large $\Gamma$, and the first-order phase transition is successfully avoided.

Thus, the ground states of the FM Potts model can be obtained by iterating the optimization under the half-hot constraint $\log_{2} Q$ times, where the computational time of each optimization is expected to be polynomial.

\subsection{Phase transition under the half-hot constraint in the limit of $Q \to \infty$}
In this subsection, we analytically confirm that the first-order phase transition can be avoided, even in the limit of $Q \to \infty$.
Note that the mean-field theory can be applied to the second term in Eq. (\ref{eq:ferro_H0_half}), and free energy can be calculated in the same manner as that described in Appendix B.
Using the Suzuki--Trotter formula and the static approximation, we obtain the free energy in the limit of $N \to \infty$ and $\beta \to \infty$ as follow:
\begin{eqnarray}
f ( \{ m_{q} \}, \{ M_{i} \} ) &=& \frac{J}{2Q} \sum_{q=1}^{Q} m_{q}^{2} - \frac{\lambda}{2N} \sum_{i=1}^{N} M_{i}^{2}  \nonumber  \\
&-& \frac{1}{\beta NQ} \sum_{q=1}^{Q} \sum_{i=1}^{N} \ln 2 \cosh \beta \Xi_{qi},
\end{eqnarray}
where
\begin{equation}
\Xi_{qi} = \sqrt{\left( Jm_{q} - \lambda M_{i} \right)^{2} + \Gamma^{2}},
\end{equation}
and $M_{i}$ is the FM order parameter for $\{ \sigma^{(z)}_{qi} | q = 1, 2, ..., Q \}$.
The order parameters, $\{ m_{q} \}$ and $\{ M_{i} \}$, are determined as the minimizer of free energy, and the saddle point equations are as follow:
\begin{equation}
m_{q} = \frac{1}{N} \sum_{i=1}^{N} \left( Jm_{q}-\lambda M_{i} \right) \frac{\tanh \beta \Xi_{qi}}{\Xi_{qi}},  \label{eq:ferro_saddle_mq}
\end{equation}
\begin{equation}
M_{i} = \frac{1}{Q} \sum_{q=1}^{Q} \left( Jm_{q}-\lambda M_{i} \right) \frac{\tanh \beta \Xi_{qi}}{\Xi_{qi}}.  \label{eq:ferro_saddle_Mi}
\end{equation}
The abovementioned simultaneous equations contain an infinite number of unknown quantities.
To obtain a physically reasonable solution to these equations, we assume the following symmetries:
\begin{itemize}
\item $M_{i}$ does not depend on $i$ ($M \equiv M_{i}$).

\item $\{ m_{q} \}$ are equally divided into $m_{q} = m^{(+)} \geq 0$ and $m_{q} = m^{(-)} \leq 0$.

\item $|m^{(+)}| = |m^{(-)}|$.
\end{itemize}
From Eqs. (\ref{eq:ferro_saddle_mq}) and (\ref{eq:ferro_saddle_Mi}), $m_{q}$ and $M_{i}$ are observed to satisfy the following equation:
\begin{equation}
\frac{1}{N} \sum_{i=1}^{N} M_{i} = \frac{1}{Q} \sum_{q=1}^{Q} m_{q}.
\end{equation}
The right-hand side of this expression vanishes under the second and third assumptions,
and the first assumption yields $M=0$.
The substitution of $M_{i}=0$ into Eq. (\ref{eq:ferro_saddle_mq}) yields
\begin{equation}
m^{(\pm)} = \frac{Jm^{(\pm)}}{\sqrt{(Jm^{(\pm)})^{2} + \Gamma^{2}}} \tanh \beta\sqrt{(Jm^{(\pm)})^{2} + \Gamma^{2}}.  \label{eq:ferro_m_pm}
\end{equation}
The right-hand side of Eq. (\ref{eq:ferro_m_pm}) is an odd function, consistent with the third assumption.
Furthermore, the saddle point equation (\ref{eq:ferro_m_pm}) is identical to that of the fully connected FM Ising model, indicating that the half-hot constraint removes the first-order phase transition during both QA and SA even when $Q \to \infty$.

\section{Annealing of the Potts glass model under the half-hot constraint}
In this section, we verify that the saddle point equations of the fully connected PG model under the half-hot constraint are identical to those of the SK model.
The encoded Ising Hamiltonian of the PG model is given as follows:
\begin{equation}
\hat{\mathcal{H}} = \hat{\mathcal{H}}_{0} + \hat{\mathcal{H}}_{q},  \label{eq:glass_Ham}
\end{equation}
\begin{equation}
\hat{\mathcal{H}}_{0} = - \sum_{i<j} J_{ij} \sum_{q=1}^{Q} \hat{\sigma}^{(z)}_{qi} \hat{\sigma}^{(z)}_{qj} + \frac{\lambda}{2Q} \sum_{i=1}^{N} \left( \sum_{q=1}^{Q} \hat{\sigma}^{(z)}_{qi} \right)^{2},  \label{eq:glass_cost}
\end{equation}
\begin{equation}
\hat{\mathcal{H}}_{\mathrm{q}} = - \Gamma \sum_{i=1}^{N} \sum_{q=1}^{Q} \hat{\sigma}^{(x)}_{qi},  \label{eq:glass_fluctuation}
\end{equation}
\begin{equation}
P \left( J_{ij} \right) = \frac{1}{J} \sqrt{ \frac{N}{2 \pi} } \exp \left[ - \frac{N}{2J^{2}} \left( J_{ij} - \frac{J_{0}}{N} \right)^{2} \right],  \label{eq:PG_Jij_distribution}
\end{equation}
where $P(J_{ij})$ represents the probability distribution of $J_{ij}$.
The Hamiltonian of the SK model is given by the following equation:
\begin{equation}
\hat{\mathcal{H}}_{\mathrm{SK}} = - \sum_{i<j} J_{ij} \hat{\sigma}^{(z)}_{i} \hat{\sigma}^{(z)}_{j},
\end{equation}
where the probability distribution of $J_{ij}$ is given by Eq. (\ref{eq:PG_Jij_distribution}).
The SK model is one of the most famous spin glass model that can be analytically investigated by the mean-field theory.
Note that the first term in Eq. (\ref{eq:glass_cost}) is the sum of the Hamiltonians of the $Q$ independent SK models, as is the case for the FM Potts model.

\subsection{Free energy and saddle point equations under the assumptions of the replica symmetric solution and the static approximation}
In this subsection, we evaluate the free energy of the PG model under the half-hot constraint and derive saddle point equations.
The free energy is given by the following equation:
\begin{equation}
- \beta [f] = \frac{1}{NQ} [ \log Z ],
\end{equation}
where the partition function $Z$ is given by the following equation:
\begin{equation}
Z = \mathrm{Tr} e^{-\beta \hat{\mathcal{H}}}.
\end{equation}
Here, the square brackets indicate that the quantity is averaged over the disorder.
Although it is extremely difficult to directly evaluate $[ \log Z ]$, we can avoid this difficulty by using the replica trick \cite{replica_trick}, which exploits the following identical equation:
\begin{equation}
[\log Z] = \lim_{n \to 0} \frac{[Z^{n}] -1}{n}.
\end{equation}
It introduces $n$ independent replicas, and the partition function of the replicated system $[Z^{n}]$ is evaluated rather than $[\log Z]$.

Using the Suzuki--Trotter formula followed by the replica trick, we obtain the free energy under the assumptions of the static approximation and replica symmetric solution in the limit of $N \to \infty$ and $Q \to \infty$ as follows (see Appendix C for detailed calculations):
\begin{eqnarray}
- \beta f = &-& \frac{\beta J_{0}}{2Q} \sum_{q} m_{q}^{2} + \frac{\beta^{2}J^{2}}{4Q} \sum_{q} \xi_{q}^{2}  \nonumber  \\
&-& \frac{\beta^{2}J^{2}}{4Q} \sum_{q} \eta_{q}^{2} + \frac{\beta \lambda}{2N} \sum_{i} M_{i}^{2}  \nonumber  \\
&+& \frac{1}{NQ} \sum_{q,i} \int Du_{qi} \ln \int Dv_{qi} 2 \cosh \beta \Xi_{qi},  \label{eq:PG_free_energy}
\end{eqnarray}
where
\begin{equation}
\Xi_{qi} = \sqrt{ H_{qi}^{2} + \Gamma^{2} },  \label{eq:glass_Xiqi}
\end{equation}
and
\begin{equation}
H_{qi} = J_{0}m_{q} - \lambda M_{i} + J \left[ \sqrt{ \xi_{q} } u_{qi} + \sqrt{ \eta_{q} - \xi_{q} } v_{qi} \right].  \label{eq:glass_Hqi_n=0}
\end{equation}
Here, $m_{q}$ and $M_{i}$ are the FM order parameters, $\xi_{q}$ represents the overlap between the different replicas and Trotter slices, $\eta_{q}$ denotes the overlap between the different Trotter slices within the same replica, $Du_{qi} \equiv du_{qi} \exp( - u_{qi}^{2} / 2 ) / \sqrt{2 \pi}$, and $Dv_{qi} \equiv dv_{qi} \exp( - v_{qi}^{2} / 2 ) / \sqrt{2 \pi}$.
The static approximation and replica symmetric solution assume that the abovementioned order parameters are independent of the Trotter slice and the replicas.
As is done for the FM Potts model, we assume the following symmetries:
\begin{itemize}
\item $M_{i}$ does not depend on $i$ ($M \equiv M_{i}$).

\item $\{ m_{q} \}$ are equally divided into $m_{q} = m^{(+)} \geq 0$ and $m_{q} = m^{(-)} \leq 0$.

\item $|m^{(+)}| = |m^{(-)}|$.
\end{itemize}
$M_{i} = 0$ can easily be derived under the assumption of the above symmetries.
Substituting $M_{i} = 0$ into Eqs. (\ref{eq:PG_free_energy}) and (\ref{eq:glass_Hqi_n=0}), we obtain the following saddle point equations:
\begin{equation}
m_{q} = \int Du_{q} \frac{ \displaystyle \int Dv_{q} \frac{H_{q}}{\Xi_{q}} \sinh \beta \Xi_{q} }{ \displaystyle \int Dv_{q} \cosh \beta \Xi_{q} },
\end{equation}
\begin{equation}
\xi_{q} = \int Du_{q} \left( \frac{ \displaystyle \int Dv_{q} \frac{H_{q}}{\Xi_{q}} \sinh \beta \Xi_{q}  }{ \displaystyle \int Dv_{q} \cosh \beta \Xi_{q} } \right)^{2},
\end{equation}
\begin{equation}
\eta_{q} = \int Du_{q} \frac{ \displaystyle \int Dv_{q} \left( \frac{H_{q}^{2}}{\Xi_{q}^{2}} \cosh \beta \Xi_{q} + \frac{\Gamma^{2}}{\beta \Xi_{q}^{3}} \sinh \beta \Xi_{q} \right)}{\displaystyle \int Dv_{q} \cosh \beta \Xi_{q} },
\end{equation}
where
\begin{equation}
\Xi_{q} = \sqrt{ H_{q}^{2} + \Gamma^{2} },
\end{equation}
and
\begin{equation}
H_{q} = \sqrt{\xi_{q}} u_{q} + \sqrt{ \eta_{q} - \xi_{q} } v_{q} + J_{0} m_{q}.
\end{equation}
These saddle point equations are identical to those of the SK model in the transverse magnetic field derived in Ref. \cite{SK_replica}, indicating that QA and SA of the PG model under the half-hot constraint undergo the same phase transitions as the SK model in the first iteration.

\subsection{Applicability of the iterative optimization to the Potts glass model}
\begin{figure}
	\includegraphics[width=8.6cm]{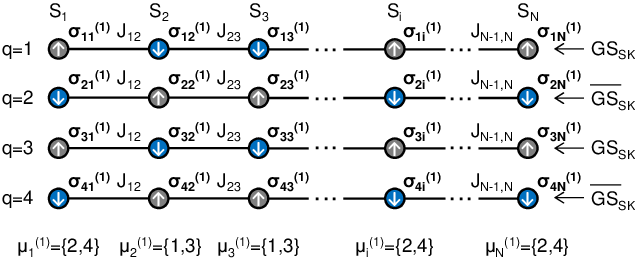}
	\caption{One of the ground states of Eq. (\ref{eq:glass_cost}).
	Vertices and edges represent Ising spins and interactions between them, respectively, and, for simplicity, only the nearest-neighbor interactions generated by the first term are shown.
	$\mathrm{GS}_{\mathrm{SK}}$ and $\overline{\mathrm{GS}}_{\mathrm{SK}}$ denote the ground states of the SK model, and the Ising spins in $\overline{\mathrm{GS}}_{\mathrm{SK}}$ are all reversed from $\mathrm{GS}_{\mathrm{SK}}$.
	Note that $\mu^{(1)}_{i}$ and $\nu^{(1)}_{i}$ depend on $i$.
	In this example, $\mu^{(1)}_{i}$ and $\nu^{(1)}_{i}$ are restricted to $\{ 1, 3 \}$ or $\{ 2, 4 \}$ because $\mathrm{GS}_{\mathrm{SK}}$ and $\overline{\mathrm{GS}}_{\mathrm{SK}}$ are alternately arranged.}
	\label{fig:glass_HH_ground}
\end{figure}

\begin{figure}
	\includegraphics[width=8.6cm]{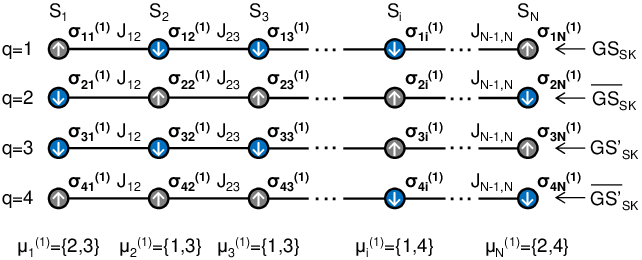}
	\caption{Another ground states of Eq. (\ref{eq:glass_cost}).
	Vertices and edges represent Ising spins and interactions between them, respectively, and, for simplicity, only the nearest-neighbor interactions generated by the first term are shown.
	$\mathrm{GS}'_{\mathrm{SK}}$ and $\overline{\mathrm{GS}'}_{\mathrm{SK}}$ are the ground states of the SK model that are different from $\mathrm{GS}_{\mathrm{SK}}$ and $\overline{\mathrm{GS}}_{\mathrm{SK}}$.
	In this case, in addition to $\{ 1, 3 \}$ or $\{ 2, 4 \}$, $\mu^{(1)}_{i}$ and $\nu^{(1)}_{i}$ are allowed to be $\{ 1, 4 \}$ or $\{ 2, 3 \}$.}
	\label{fig:glass_HH_ground2}
\end{figure}

Although the ground states of the FM Potts model can be obtained by the proposed method, whether the proposed method can retrieve the ground states of the PG model is unclear.
In this subsection, we show that the cost function in the second iteration is different from that in the first iteration for the PG model.
As is the case for the FM Potts model, the first term in Eq. (\ref{eq:glass_cost}) is the sum of the Hamiltonians of the $Q$ independent SK models, and the ground states of Eq. (\ref{eq:glass_cost}) comprise those of the SK model.
However, the situation is more complicated because the ground states of the SK model might be degenerated that originates from frustrations.
One of the simplest ground states of Eq. (\ref{eq:glass_cost}) for $Q=4$ is depicted in Fig. \ref{fig:glass_HH_ground}, where $\mathrm{GS}_{\mathrm{SK}}$ and $\overline{\mathrm{GS}}_{\mathrm{SK}}$ denote the ground states of the SK model, and the spins in $\overline{\mathrm{GS}}_{\mathrm{SK}}$ are all reversed from $\mathrm{GS}_{\mathrm{SK}}$.
The spin configuration in Fig. \ref{fig:glass_HH_ground} minimizes the first term of Eq. (\ref{eq:glass_cost}) and satisfies the half-hot constraint, indicating that the second term is also minimized.
In addition, if there exists a ground state of the SK model $\mathrm{GS}'_{\mathrm{SK}}$ that is different from $\mathrm{GS}_{\mathrm{SK}}$ and $\overline{\mathrm{GS}}_{\mathrm{SK}}$, the spin configuration shown in Fig. \ref{fig:glass_HH_ground2} also minimizes the cost function [Eq. (\ref{eq:glass_cost})].
Unlike the FM Potts model, $\bm{\mu}^{(1)}_{i}$ and $\bm{\nu}^{(1)}_{i}$ depend on $i$ in both the ground states.

\begin{figure}
	\includegraphics[width=8.6cm]{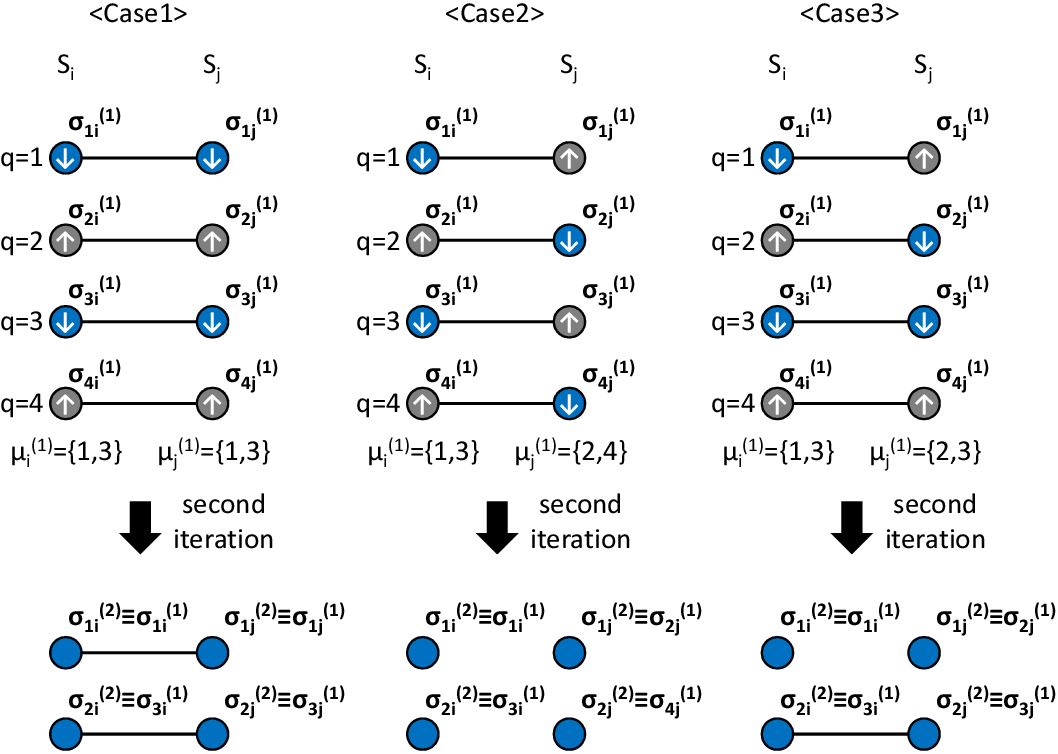}
	\caption{Local interactions in the second iteration for $Q=4$.
	Depending on $\bm{\mu}^{(1)}_{i}$ and $\bm{\mu}^{(1)}_{j}$, the local interaction in the second iteration will change from that in the first iteration.}
	\label{fig:glass_local_interaction}
\end{figure}

The local interactions in the second iteration are shown in Fig. \ref{fig:glass_local_interaction}, and there are essentially three cases.
\begin{enumerate}
\item
The same components are selected for the adjacent Potts spins in the first iteration.
In this case, except for the number of components, the local interactions in the first and second iterations are identical.

\item
Different components are selected for the adjacent Potts spins in the first iteration.
In this case, the Potts spins are independent in the second iteration.

\item
One of the selected components is the same with the adjacent Potts spins.
In this case, only one interaction $J_{ij}$ exists between $\bm{\sigma}^{(2)}_{i}$ and $\bm{\sigma}^{(2)}_{j}$ in the second iteration.
\end{enumerate} 
Thus, depending on the ground states of the SK model obtained by the optimization in the first iteration, the cost function in the second iteration differs from that in the first iteration.
Although the proposed method should be applicable to the PG model, it is not verified whether any combinations of ground states of the SK model ultimately converge to that of the PG model.
The validity of the proposed method for the PG model must be investigated in a future study.

\section{Discussion and conclusion}
We analytically investigated the order of phase transitions during QA and SA in the fully connected FM Potts model with one-hot encoding.
As with SA, QA undergoes the first-order phase transition when $Q > 2$.
It is conjectured that the computational time of QA exponentially increases with the system size $N$, and QA does not exhibit exponential acceleration for obtaining the ground state of the FM Potts model.

To avoid the first-order phase transition, we proposed the iterative optimization under the half-hot constraint.
The proposed method is applicable to both QA and SA, and it is capable of retrieving the ground states of the FM Potts model.
The first-order phase transition is caused by the longitudinal magnetic field in the penalty term of the one-hot constraint, which positively biases the order parameters $m^{(\pm)} > 0$ when $\Gamma$ is large.
In contrast, the half-hot constraint cancels the longitudinal magnetic field in the penalty term.
Consequently, as $\Gamma$ decreases, $m^{(\pm)}$ continuously change from $m^{(\pm)}=0$ to $m^{(\pm)} \gtrless 0$.
We confirmed that the saddle point equation is identical to that of the fully connected FM Ising model in the limit of $Q \to \infty$.
Therefore, the first-order phase transition is avoided even when $Q \to \infty$, which indicates that the ground states of the FM Potts model can be obtained by iterating the optimization under the half-hot constraint $\log_{2} Q$ times,
where the computational time of each optimization is expected to be polynomial.

We further investigated the optimization of the fully connected PG model under the half-hot constraint.
As observed in the FM Potts model, the saddle point equations of the PG model under the half-hot constraint are identical to those of the SK model in the limit of $Q \to \infty$.
This result indicates that the QA and SA of the PG model under the half-hot constraint undergo the same phase transitions as those of the SK model.
The ground states in the first iteration comprises those of the SK model.
The Hamiltonian in the second iteration strongly depends on the ground states of the SK model obtained in the first iteration.
Although the proposed method is expected to be applicable to the PG model, whether any combinations of the ground states of the SK model ultimately lead to those of the PG model is not clarified and should be assessed in a future study.

\begin{acknowledgments}
The authors are very grateful to Tadashi Kadowaki and Masamichi J. Miyama for their fruitful discussions.
M. O. is grateful for the financial support provided by JSPS KAKENHI 19H01095 and 16H04382, Next Generation High-Performance Computing Infrastructures and Applications R\&D Programby MEXT.
K. T. was partially supported by a Grant-in-Aid for Scientific Research from the Japan Society.
K. T. was partially supported by JSPS KAKENHI No. 18H03303
\end{acknowledgments}

\appendix
\section{Free energy of the ferromagnetic Potts model under the one-hot constraint}
We derive the free energy [Eqs. (\ref{eq:ferro_OH_free1}) and (\ref{eq:ferro_OH_free2})] of the FM Potts model whose Hamiltonian is defined in Eqs. (\ref{eq:ferro_H0_Hq}), (\ref{eq:ferro_one-hot_rep}), and (\ref{eq:ferro_one-hot_fluctuation}).
The Suzuki--Trotter formula and static approximation enable us to calculate the partition function.

The partition function $Z$ is given by the following equation:
\begin{equation}
Z = \mathrm{Tr} e^{-\beta \hat{\mathcal{H}}_{0} - \beta \hat{\mathcal{H}}_{\mathrm{q}}},
\end{equation}
where $\hat{\mathcal{H}}_{0}$ and $\hat{\mathcal{H}}_{\mathrm{q}}$ are defined in Eqs. (\ref{eq:ferro_one-hot_rep}) and (\ref{eq:ferro_one-hot_fluctuation}).
Using the Suzuki--Trotter formula, the partition function is rewritten as follows:
\begin{equation}
Z = \lim_{K \to \infty} \mathrm{Tr} \left[ \left( - \frac{\beta}{K} \hat{\mathcal{H}}_{0} \right) \exp \left( - \frac{\beta}{K} \hat{\mathcal{H}}_{\mathrm{q}} \right) \right]^{K}.
\end{equation}
By introducing the closure relations:
\begin{equation}
\hat{1}(\kappa) = \sum_{\bm{\sigma}^{(z)}(\kappa)} \ket{\bm{\sigma}^{(z)}(\kappa)} \bra{\bm{\sigma}^{(z)}(\kappa)},  \label{eq:Appendix_A_closure}
\end{equation}
and substituting Eq. (\ref{eq:ferro_one-hot_rep}) into the resulting expression, we obtain
\begin{widetext}
\begin{eqnarray}
Z = \lim_{K \to \infty} \prod_{\kappa=1}^{K} &&\sum_{\bm{\sigma}^{(z)}(\kappa)} \Braket{\bm{\sigma}^{(z)}(\kappa)|\exp \left( - \frac{\beta}{K} \hat{\mathcal{H}}_{\mathrm{q}} \right)|\bm{\sigma}^{(z)}(\kappa+1)}  \nonumber  \\
&&\times \exp \left\{ \frac{\beta J}{2NK} \sum_{q} \left( \sum_{i} \sigma^{(z)}_{qi}(\kappa) \right)^{2} - \frac{\beta \lambda}{2QK} \sum_{i} \left[ \left( \sum_{q} \sigma^{(z)}_{qi}(\kappa) \right)^{2} - 2(Q-2) \sum_{q} \sigma^{(z)}_{qi} (\kappa) \right] \right\},  \label{eq:Appendix_A_ST}
\end{eqnarray}
\end{widetext}
where $\kappa$ represents the Trotter slice, $\sigma^{(z)}_{qi}(\kappa)$ is the Ising spin assigned to the component $q$ of $S_{i}$ in the Trotter slice $\kappa$, $\bm{\sigma}^{(z)}(\kappa)$ denotes Ising spins in the Trotter slice $\kappa$, namely $\{ \sigma^{(z)}_{qi} (\kappa) | i=1, 2, ..., N, q = 1, 2, ..., Q\}$,
and $\sum_{\bm{\sigma}^{(z)}(\kappa)}$ represents the summation over all spin configurations of $\bm{\sigma}^{(z)}(\kappa)$.
Subsequently, we linearize the spin-product term using the Hubbard--Stratonovich transformation \cite{Hubbard_transform} as follows:
\begin{widetext}
\begin{eqnarray}
\exp \left[ \frac{\beta J}{2NK} \sum_{q} \left( \sum_{i} \sigma^{(z)}_{qi}(\kappa) \right)^{2} \right] \propto \int d \bm{m}(\kappa) \exp \left[ - \frac{\beta JN}{2K} \sum_{q} (m_{q}(\kappa))^{2} + \frac{\beta J}{K} \sum_{q} m_{q}(\kappa) \sum_{i} \sigma^{(z)}_{qi}(\kappa) \right],  \label{eq:Appendix_A_gauss}
\end{eqnarray}
\end{widetext}
where $d \bm{m}(\kappa) \equiv \prod_{q} dm_{q}(\kappa)$, and $m_{q}(\kappa)$ is the FM order parameter of $\{ \sigma^{(z)}_{qi}(\kappa) | i=1, 2, ..., N \}$.
Substitution of Eq. (\ref{eq:Appendix_A_gauss}) into Eq. (\ref{eq:Appendix_A_ST}) and inverse operation of the closure relation [Eq. (\ref{eq:Appendix_A_closure})] yields
\begin{eqnarray}
Z &\propto& \lim_{K \to \infty} \int d \bm{m} \exp \left[ - \frac{\beta JN}{2K} \sum_{q, \kappa} (m_{q}(\kappa))^{2} \right]   \nonumber  \\
&\times& \mathrm{Tr} \left[ \prod_{k} \exp\left( - \frac{\beta}{K} \hat{\mathcal{H}}^{(\kappa)}_{0} \right)  \exp \left( - \frac{\beta}{K} \hat{\mathcal{H}}_{\mathrm{q}} \right) \right],
\end{eqnarray}
where
\begin{eqnarray}
\hat{\mathcal{H}}^{(\kappa)}_{0} \equiv \frac{\lambda}{2Q}&& \sum_{i} \left( \sum_{q} \hat{\sigma}^{(z)}_{qi} \right)^{2}  \nonumber  \\
&&- \sum_{q,i} \left( Jm_{q}(\kappa) + \frac{Q-2}{Q} \lambda \right) \hat{\sigma}^{(z)}_{qi},
\end{eqnarray}
and $d \bm{m} \equiv \prod_{\kappa, q} dm_{q}(\kappa)$.
By applying the static approximation, which neglects the $\kappa$-dependence of $m_{q}(\kappa)$, followed by the inverse operation of the Suzuki--Trotter formula, we obtain
\begin{equation}
Z \propto \int d \bm{m} e^{- \beta N f(\{m_{q}\})},  \label{eq:Appendix_A_final_Z}
\end{equation}
\begin{equation}
f(\{m_{q}\}) = \frac{J}{2} \sum_{q} m_{q}^{2} - \frac{1}{\beta} \log \mathrm{Tr} e^{-\beta \hat{\mathcal{H}}^{(\mathrm{eff})}},
\end{equation}
where
\begin{eqnarray}
\hat{\mathcal{H}}^{(\mathrm{eff})} && ( \{ m_{q} \} ) = \frac{\lambda}{2Q} \left( \sum_{q} \hat{\sigma}^{(z)}_{q} \right)^{2}  \nonumber  \\
- \sum_{q}&& \left( J m_{q} + \frac{Q-2}{Q} \lambda \right) \hat{\sigma}^{(z)}_{q} - \Gamma \sum_{q} \hat{\sigma}^{(x)}_{q}.
\end{eqnarray}
The exponent of Eq. (\ref{eq:Appendix_A_final_Z}) is proportional to $N$, and the integral of $\bm{m}$ can be evaluated by the saddle point method in the limit of $N \to \infty$.
In the limit of $\beta \to \infty$, only the lowest eigenvalue $\varepsilon^{(\mathrm{eff})}_{\mathrm{min}} ( \{ m_{q} \} )$ of $\hat{\mathcal{H}}^{(\mathrm{eff})} ( \{ m_{q} \} )$ contributes to free energy.
Therefore, the free energy at $\beta \to \infty$ is given by Eqs. (\ref{eq:ferro_OH_free1}) and (\ref{eq:ferro_OH_free2}).

\section{Cost function of the iterative optimization method}
We derive a general formula of the cost function in the $(k+1)$th iteration from the optimization result in the $k$th iteration.
The optimization problem in the second iteration for the FM Potts model [Eq. (\ref{eq:ferro_second_iteration})] can be obtained using this formula.

The general form of the optimization problem in the $k$th iteration is given by the following equation:
\begin{eqnarray}
\underset{\bm{\sigma}^{(k)}}{\mathrm{argmin}} \left[ - \sum_{i<j} \sum_{q,q'=1}^{Q^{(k)}} J^{(k)}_{ij}(q,q') \sigma^{(k)}_{qi} \sigma^{(k)}_{q'j} \right.&&  \nonumber  \\
\left. - \sum_{i} \sum_{q=1}^{Q^{(k)}} h^{(k)}_{i}(q) \sigma^{(k)}_{qi} \right] \ \mathrm{s.t.} \ \sum_{q=1}^{Q^{(k)}} \sigma^{(k)}_{qi}&& = 0,  \label{eq:Appendix_B_kth_opt}
\end{eqnarray}
where $\sigma^{(k)}_{qi}$ is the spin variable in the $k$th iteration, $J^{(k)}_{ij}(q,q')$ is the interaction between $\sigma^{(k)}_{qi}$ and $\sigma^{(k)}_{q'j}$, $h^{(k)}_{i}(q)$ is the longitudinal magnetic field applied to $\sigma^{(k)}_{qi}$, and $Q^{(k)} = Q / 2^{k-1}$ is the number of components in the $k$th iteration.
The optimization result in the $k$th iteration is determined by $\bm{\mu}^{(k)}$ and $\bm{\nu}^{(k)}$.
Note that $\{ \sigma^{(k)}_{qi} = -1 | q \in \bm{\mu}^{(k)}_{i} \}$ and $\{ \sigma^{(k)}_{qi} = +1 | q \in \bm{\nu}^{(k)}_{i} \}$.
The cost function in the $(k+1)$th iteration can be derived by fixing $\{ \sigma^{(k)}_{qi} | q \in \bm{\nu}^{(k)}_{i} \}$ to $+1$ and replacing $\sigma^{(k)}_{\mu^{(k)}_{qi},i}$ to $\sigma^{(k+1)}_{qi}$.
By rewriting the summation over $q$ and $q'$ in Eq. (\ref{eq:Appendix_B_kth_opt}) as follows:
\begin{eqnarray}
\sum_{q} &=& \sum_{q \in \bm{\mu}^{(k)}_{i}} + \sum_{q \in \bm{\nu}^{(k)}_{i}},  \\
\sum_{q'} &=& \sum_{q' \in \bm{\mu}^{(k)}_{j}} + \sum_{q' \in \bm{\nu}^{(k)}_{j}},
\end{eqnarray}
we obtain
\begin{eqnarray}
- \sum_{i<j} \sum_{q,q'} J^{(k+1)}_{ij}&&(q,q') \sigma^{(k+1)}_{qi} \sigma^{(k+1)}_{q'j}  \nonumber  \\
-&& \sum_{q,i} h^{(k+1)}_{i}(q) \sigma^{(k+1)}_{qi} + \mathrm{const.},
\end{eqnarray}
where
\begin{equation}
J^{(k+1)}_{ij}(q,q') = J^{(k)}_{ij} ( \mu^{(k)}_{qi}, \mu^{(k)}_{q'j} ),  \label{eq:Appendix_B_Jij_k+1}
\end{equation}
\begin{equation}
h^{(k+1)}_{i}(q) = \sum_{j \neq i} \sum_{q'} J^{(k)}_{ij}( \mu^{(k)}_{qi}, \nu^{(k)}_{q'j} ) + h^{(k)}_{i}(\mu^{(k)}_{qi}).  \label{eq:Appendix_B_hi_k+1}
\end{equation}
In the FM Potts model, the interaction and longitudinal magnetic field of the first iteration are given by the following equation:
\begin{equation}
J^{(1)}_{ij}(q,q') = \frac{J}{N} \delta (q,q'),  \label{eq:Appendix_B_ferro_Jij}
\end{equation}
\begin{equation}
h^{(1)}_{i}(q) = 0,  \label{eq:Appendix_B_ferro_hi}
\end{equation}
where $\delta$ is the Kronecker delta function.
By substituting Eqs. (\ref{eq:Appendix_B_ferro_Jij}) and (\ref{eq:Appendix_B_ferro_hi}) into Eqs. (\ref{eq:Appendix_B_Jij_k+1}) and (\ref{eq:Appendix_B_hi_k+1}), we obtain
\begin{equation}
J^{(2)}_{ij}(q,q') = \frac{J}{N} \delta ( \mu^{(1)}_{qi}, \mu^{(1)}_{q'j} ),
\end{equation}
\begin{equation}
h^{(2)}_{i}(q) = \frac{J}{N} \sum_{j \neq i} \sum_{q'} \delta( \mu^{(1)}_{qi}, \nu^{(1)}_{q'j} ),
\end{equation}
yielding Eq. (\ref{eq:ferro_second_iteration}).

\section{Free energy of the Potts glass model under the half-hot constraint}
We derive the free energy [Eqs. (\ref{eq:PG_free_energy}), (\ref{eq:glass_Xiqi}), and (\ref{eq:glass_Hqi_n=0})] of the PG model under the half-hot constraint whose Hamiltonian is defined in Eqs. (\ref{eq:glass_Ham}), (\ref{eq:glass_cost}), (\ref{eq:glass_fluctuation}), and (\ref{eq:PG_Jij_distribution}).
According to the replica trick, the partition function of the $n$ replicated system needs to be calculated.

Applying the Suzuki--Trotter formula, we obtain the following partition function $[Z^{n}]$:
\begin{widetext}
\begin{eqnarray}
[Z^{n}] = \lim_{K \to \infty} e^{nNQC} \mathrm{Tr} \exp \left[ - \frac{\beta \lambda}{2KQ} \sum_{\kappa,\alpha} \sum_{i} \left( \sum_{q} \sigma^{(\kappa,\alpha)}_{qi} \right)^{2} + \frac{1}{2} \ln \coth \left( \frac{\beta \Gamma}{K} \right) \sum_{q,i} \sum_{\kappa,\alpha} \sigma^{(\kappa,\alpha)}_{qi} \sigma^{(\kappa+1,\alpha)}_{qi} \right]  \nonumber  \\
	\times \prod_{i<j} \int dJ_{ij} P(J_{ij}) \exp \left( \frac{\beta}{K} J_{ij} \sum_{q} \sum_{\kappa,\alpha} \sigma^{(\kappa,\alpha)}_{qi} \sigma^{(\kappa,\alpha)}_{qj} \right),
\end{eqnarray}
where
\begin{equation}
C = \frac{K}{2} \ln \sinh \left( \frac{\beta \Gamma}{K} \right) \cosh \left( \frac{\beta \Gamma}{K} \right),
\end{equation}
and $\sigma^{(\kappa,\alpha)}_{qi}$ represents the $z$ spin assigned to the component $q$ of $S_{i}$ in the Trotter slice $\kappa$ and replica $\alpha$.
The exponent of the integrand contains the summation over $q$ in addition to $\kappa$ and $\alpha$.
Therefore, to explicitly formulate the free energy, assumptions of $q$-dependence of the order parameters are required.
The integrals of $J_{ij}$ can be easily calculated as follows:
\begin{equation}
\int dJ_{ij} P(J_{ij}) \exp \left( \frac{\beta}{K} J_{ij} \sum_{q} \sum_{\kappa, \alpha} \sigma^{(\kappa,\alpha)}_{qi} \sigma^{(\kappa,\alpha)}_{qj} \right) = \exp \left[ \frac{\beta J_{0}}{NK} \sum_{q} \sum_{\kappa,\alpha} \sigma^{(\kappa,\alpha)}_{qi} \sigma^{(\kappa,\alpha)}_{qj} + \frac{\beta^{2} J^{2}}{2NK^{2}} \left( \sum_{q} \sum_{\kappa,\alpha} \sigma^{(\kappa,\alpha)}_{qi} \sigma^{(\kappa,\alpha)}_{qj} \right)^{2} \right].  \label{eq:A_integral_Jij}
\end{equation}
The first and second terms of the exponent in Eq. (\ref{eq:A_integral_Jij}) are rewritten as follows:
\begin{equation}
\sum_{i<j} \sum_{q} \sum_{\kappa,\alpha} \sigma^{(\kappa,\alpha)}_{qi} \sigma^{(\kappa,\alpha)}_{qj} = \frac{1}{2} \sum_{q} \sum_{\kappa,\alpha} \left[ \left( \sum_{i} \sigma^{(\kappa,\alpha)}_{qi} \right)^{2} - N \right],
\end{equation}
\begin{eqnarray}
\sum_{i<j} \left( \sum_{q} \sum_{\kappa,\alpha} \sigma^{(\kappa,\alpha)}_{qi} \sigma^{(\kappa,\alpha)}_{qj} \right)^{2} &=& \sum_{\kappa,\kappa'} \sum_{\alpha < \alpha'} \left[ \sum_{q \neq q'} \left( \sum_{i} \sigma^{(\kappa,\alpha)}_{qi} \sigma^{(\kappa',\alpha')}_{q'i} \right)^{2} + \sum_{q} \left( \sum_{i} \sigma^{(\kappa,\alpha)}_{qi} \sigma^{(\kappa',\alpha')}_{qi} \right)^{2} - NQ^{2} \right]  \nonumber  \\
&+& \frac{1}{2} \sum_{\kappa,\kappa'} \sum_{\alpha} \left[ \sum_{q \neq q'} \left( \sum_{i} \sigma^{(\kappa,\alpha)}_{qi} \sigma^{(\kappa',\alpha)}_{q'i} \right)^{2} + \sum_{q} \left( \sum_{i} \sigma^{(\kappa,\alpha)}_{qi} \sigma^{(\kappa',\alpha)}_{qi} \right)^{2} - NQ^{2} \right].
\end{eqnarray}
By substituting the above equations into Eq. (\ref{eq:A_integral_Jij}), we obtain
\begin{eqnarray}
\prod_{i<j} \int dJ_{ij} P(J_{ij}) \exp \left( \frac{\beta}{K} J_{ij} \sum_{q} \sum_{\kappa,\alpha} \sigma^{(\kappa,\alpha)}_{qi} \sigma^{(\kappa,\alpha)}_{qj} \right) &&\approx \exp \left[ \frac{\beta J_{0}}{2NK} \sum_{q} \sum_{\kappa,\alpha} \left( \sum_{i} \sigma^{(\kappa,\alpha)}_{qi} \right)^{2} \right.  \nonumber  \\
+ \frac{\beta^{2}J^{2}}{2NK^{2}} \sum_{\kappa,\kappa'} \sum_{\alpha < \alpha'} \sum_{q} \left( \sum_{i} \sigma^{(\kappa,\alpha)}_{qi} \sigma^{(\kappa',\alpha')}_{qi} \right)^{2} +&& \frac{\beta^{2}J^{2}}{2NK^{2}} \sum_{\kappa,\kappa'} \sum_{\alpha < \alpha'} \sum_{q \neq q'} \left( \sum_{i} \sigma^{(\kappa,\alpha)}_{qi} \sigma^{(\kappa',\alpha')}_{q'i} \right)^{2}  \nonumber  \\
+ \frac{\beta^{2}J^{2}}{4NK^{2}} \sum_{\kappa,\kappa'} \sum_{\alpha} \sum_{q} \left( \sum_{i} \sigma^{(\kappa,\alpha)}_{qi} \sigma^{(\kappa',\alpha)}_{qi} \right)^{2}&& + \left. \frac{\beta^{2}J^{2}}{4NK^{2}} \sum_{\kappa,\kappa'} \sum_{\alpha} \sum_{q \neq q'} \left( \sum_{i} \sigma^{(\kappa,\alpha)}_{qi} \sigma^{(\kappa',\alpha)}_{q'i} \right)^{2} \right].
\end{eqnarray}
Subsequently, by applying the Hubbard--Stratonovich transformation for each term of the exponent, we obtain the free energy as follows:
\begin{eqnarray}
- \beta [f] = \lim_{n \to 0} \lim_{K \to \infty} \left\{ - \frac{\beta J_{0}}{2QK n} \sum_{q,\kappa,\alpha} (m^{(\alpha)}_{q\kappa})^{2} \right. - \frac{\beta^{2}J^{2}}{2QK^{2}n} \sum_{q} \sum_{\kappa,\kappa'} \sum_{\alpha < \alpha'}&& (\xi^{(\alpha \alpha')}_{q,\kappa \kappa'})^{2}   \nonumber  \\
- \frac{\beta^{2}J^{2}}{2QK^{2}n} \sum_{q \neq q'} \sum_{\kappa,\kappa'} \sum_{\alpha < \alpha'} (\theta^{(\alpha \alpha')}_{qq',\kappa \kappa'})^{2} - \frac{\beta^{2}J^{2}}{4QK^{2}n} \sum_{q} \sum_{\kappa,\kappa'} \sum_{\alpha} (\eta^{(\alpha)}_{q,\kappa \kappa'})^{2} &&- \frac{\beta^{2}J^{2}}{4QK^{2}n} \sum_{q \neq q'} \sum_{\kappa,\kappa'} \sum_{\alpha} (\varphi^{(\alpha)}_{qq',\kappa \kappa'})^{2}  \nonumber  \\
+ \frac{\beta \lambda}{2NK n}&& \sum_{i,\kappa,\alpha} (M^{(\alpha)}_{i \kappa})^{2} + \left. \frac{1}{NQn} \sum_{i} \ln \mathrm{Tr} e^{L_{i}} + C \right\},
\end{eqnarray}
where
\begin{eqnarray}
L_{i} = \frac{\beta^{2}J^{2}}{K^{2}} \sum_{q} \sum_{\kappa,\kappa'} \sum_{\alpha < \alpha'} \xi^{(\alpha \alpha')}_{q,\kappa \kappa'} \sigma^{(\kappa,\alpha)}_{qi}&& \sigma^{(\kappa',\alpha')}_{qi} + \frac{\beta^{2}J^{2}}{K^{2}} \sum_{q \neq q'} \sum_{\kappa,\kappa'} \sum_{\alpha < \alpha'} \theta^{(\alpha \alpha')}_{qq',\kappa \kappa'} \sigma^{(\kappa,\alpha)}_{qi} \sigma^{(\kappa',\alpha')}_{q'i}  \nonumber  \\
 + \frac{\beta^{2}J^{2}}{2K^{2}} \sum_{q} \sum_{\kappa,\kappa'} \sum_{\alpha}&& \eta^{(\alpha)}_{q,\kappa \kappa'} \sigma^{(\kappa,\alpha)}_{qi} \sigma^{(\kappa',\alpha)}_{qi} + \frac{\beta^{2}J^{2}}{2K^{2}} \sum_{q \neq q'} \sum_{\kappa,\kappa'} \sum_{\alpha} \varphi^{(\alpha)}_{qq',\kappa \kappa'} \sigma^{(\kappa,\alpha)}_{qi} \sigma^{(\kappa',\alpha)}_{q'i}  \nonumber  \\
+ \frac{\beta}{K}&& \sum_{q,\kappa,\alpha} \left( J_{0} m^{(\alpha)}_{q \kappa} - \lambda M^{(\alpha)}_{i \kappa} \right) \sigma^{(\kappa,\alpha)}_{qi} + \frac{1}{2} \ln \coth \left( \frac{\beta \Gamma}{K} \right) \sum_{q,\kappa,\alpha} \sigma^{(\kappa,\alpha)}_{qi} \sigma^{(\kappa+1,\alpha)}_{qi}.  \label{eq:A_glass_Heff}
\end{eqnarray}
\end{widetext}
The saddle point equations are given as follows:
\begin{eqnarray}
m^{(\alpha)}_{q \kappa} &=& \frac{1}{N} \sum_{i}^{N} \left< \sigma^{(\kappa,\alpha)}_{qi} \right>_{L_{i}},  \\
\xi^{(\alpha \alpha')}_{q,\kappa \kappa'} &=& \frac{1}{N} \sum_{i} \left< \sigma^{(\kappa,\alpha)}_{qi} \sigma^{(\kappa',\alpha')}_{qi} \right>_{L_{i}},  \\
\theta^{(\alpha \alpha')}_{qq',\kappa \kappa'} &=& \frac{1}{N} \sum_{i} \left< \sigma^{(\kappa,\alpha)}_{qi} \sigma^{(\kappa',\alpha')}_{q'i} \right>_{L_{i}},  \\
\eta^{(\alpha)}_{q,\kappa \kappa'} &=& \frac{1}{N} \sum_{i} \left< \sigma^{(\kappa,\alpha)}_{qi} \sigma^{(\kappa',\alpha)}_{qi} \right>_{L_{i}},  \\
\varphi^{(\alpha)}_{qq',\kappa \kappa'} &=& \frac{1}{N} \sum_{i} \left< \sigma^{(\kappa,\alpha)}_{qi} \sigma^{(\kappa',\alpha)}_{q'i} \right>_{L_{i}},  \\
M^{(\alpha)}_{i\kappa} &=& \left< \frac{1}{Q} \sum_{q} \sigma^{(\kappa,\alpha)}_{qi} \right>_{L_{i}}.
\end{eqnarray}

Next, we apply the static approximation and assume the replica symmetric solution as shown below:
\newpage
\begin{eqnarray}
m_{q} &\equiv& m^{(\alpha)}_{q \kappa},  \label{eq:A_sym1}  \\
\xi_{q} &\equiv& \xi^{(\alpha \alpha')}_{q,\kappa \kappa'},  \label{eq:A_sym2}  \\
\theta_{qq'} &\equiv& \theta^{(\alpha \alpha')}_{qq',\kappa \kappa'},  \label{eq:A_sym3}  \\
\eta_{q} &\equiv& \eta^{(\alpha)}_{q,\kappa \kappa'},  \label{eq:A_sym4}  \\
\varphi_{qq'} &\equiv& \varphi^{(\alpha)}_{qq',\kappa \kappa'},  \label{eq:A_sym5}  \\
M_{i} &\equiv& M^{(\alpha)}_{i \kappa}.  \label{eq:A_sym6}
\end{eqnarray}
As $q = 1, 2, ..., Q$ are equally treated in Eqs. (\ref{eq:glass_Ham}), (\ref{eq:glass_cost}), and (\ref{eq:glass_fluctuation}),
we further assume that $\theta_{qq'}$ and $\varphi_{qq'}$ are independent of $(q, q')$.
\begin{eqnarray}
\theta &\equiv& \theta_{qq'},  \label{eq:A_sym7}  \\
\varphi &\equiv& \varphi_{qq'}.  \label{eq:A_sym8}
\end{eqnarray}
The resulting expression of $L_{i}$ is given by the following equation:
\begin{widetext}
\begin{eqnarray}
L_{i} &=& \frac{\beta^{2}J^{2}}{2K^{2}} \sum_{q} \left( \xi_{q} - \theta \right) \left( \sum_{\kappa,\alpha} \sigma^{(\kappa,\alpha)}_{qi} \right)^{2} + \frac{\beta^{2}J^{2}}{2K^{2}} \sum_{q} \left[ \left( \eta_{q} - \xi_{q} \right) - \left( \varphi - \theta \right) \right] \sum_{\alpha} \left( \sum_{\kappa} \sigma^{(\kappa,\alpha)}_{qi} \right)^{2} + \frac{\beta^{2}J^{2}}{2K^{2}} \theta \left( \sum_{q,\kappa,\alpha} \sigma^{(\kappa,\alpha)}_{qi} \right)^{2}  \nonumber  \\
&+& \frac{\beta^{2}J^{2}}{2K^{2}} \left( \varphi - \theta \right) \sum_{\alpha} \left( \sum_{q,\kappa} \sigma^{(\kappa,\alpha)}_{qi} \right)^{2} + \frac{\beta}{K} \sum_{q,\kappa,\alpha} \left( J_{0}m_{q} - \lambda M_{i} \right) \sigma^{(\kappa,\alpha)}_{qi} + \frac{1}{2} \ln \coth \left( \frac{\beta \Gamma}{K} \right) \sum_{q,\kappa,\alpha} \sigma^{(\kappa,\alpha)}_{qi} \sigma^{(\kappa+1,\alpha)}_{qi}.  \label{eq:A_Li_sym}
\end{eqnarray}
Applying the Hubbard--Stratonovich transformation to the first and second terms in $L_{i}$, we obtain
\begin{equation}
\exp \left[ \frac{\beta^{2}J^{2}}{2K^{2}} \sum_{q} \left( \xi_{q} - \theta \right) \left( \sum_{\kappa,\alpha} \sigma^{(\kappa,\alpha)}_{qi} \right)^{2} \right] = \int D \bm{u}_{i} \exp \left[ \frac{\beta J}{K} \sum_{q} \sqrt{ \xi_{q} - \theta } \left( \sum_{\kappa,\alpha} \sigma^{(\kappa,\alpha)}_{qi} \right) u_{qi} \right],  \label{eq:A_Li_sym_1st}
\end{equation}
and
\begin{eqnarray}
\exp&& \left\{ \frac{\beta^{2}J^{2}}{2K^{2}} \sum_{q} \left[ \left( \eta_{q} - \xi_{q} \right) - \left( \varphi - \theta \right) \right] \sum_{\alpha} \left( \sum_{\kappa} \sigma^{(\kappa,\alpha)}_{qi} \right)^{2} \right\}  \nonumber  \\
&&\ \ \ \ \ \ \ \ \ \ = \int D \bm{v}_{i} \exp \left[ \frac{\beta J}{K} \sum_{q} \sqrt{ \left( \eta_{q} - \xi_{q} \right) - \left( \varphi - \theta \right) } \sum_{\alpha} \left( \sum_{\kappa} \sigma^{(\kappa,\alpha)}_{qi} \right) v^{(\alpha)}_{qi} \right],   \label{eq:A_Li_sym_2nd}
\end{eqnarray}
where
\begin{equation}
D \bm{u}_{i} \equiv \prod_{q} \frac{du_{qi}}{\sqrt{2\pi}} \exp \left( -\frac{u_{qi}^{2}}{2} \right), \ \ \ D \bm{v}_{i} \equiv \prod_{q,\alpha} \frac{dv^{(\alpha)}_{qi}}{\sqrt{2\pi}} \exp \left[ - \frac{(v^{(\alpha)}_{qi})^{2}}{2} \right].
\end{equation}
Using the $\delta$ function and its Fourier transformation to the third and fourth terms in $L_{i}$ yields
\begin{eqnarray}
\exp&& \left[ \frac{\beta^{2}J^{2}}{2K^{2}} \theta \left( \sum_{q,\kappa,\alpha} \sigma^{(\kappa,\alpha)}_{qi} \right)^{2} + \frac{\beta^{2}J^{2}}{2K^{2}} \left( \varphi - \theta \right) \sum_{\alpha} \left( \sum_{q,\kappa} \sigma^{(\kappa,\alpha)}_{qi} \right)^{2} \right]  \nonumber  \\
&&= \int d \bm{W}_{i} \int d \bm{\omega}_{i}  \exp \left( \frac{\beta^{2}J^{2}Q}{K^{2}} \sum_{\kappa,\alpha} \omega^{(\alpha)}_{i\kappa} \sum_{q} \sigma^{(\kappa,\alpha)}_{qi} \right)  \nonumber  \\
&&\times \exp \left[ \frac{\beta^{2}J^{2}Q^{2}}{2K^{2}} \theta \left( \sum_{\kappa,\alpha} W^{(\alpha)}_{i \kappa} \right)^{2} + \frac{\beta^{2}J^{2}Q^{2}}{2K^{2}} \left( \varphi - \theta \right) \sum_{\alpha} \left( \sum_{\kappa} W^{(\alpha)}_{i \kappa} \right)^{2} - \frac{\beta^{2}J^{2}Q^{2}}{K^{2}} \sum_{\kappa,\alpha} \omega^{(\alpha)}_{i \kappa} W^{(\alpha)}_{i \kappa} \right].  \label{eq:A_Li_sym_3rd}
\end{eqnarray}
where $d \bm{W}_{i} \equiv \prod_{\kappa,\alpha} dW^{(\alpha)}_{i \kappa}$, and $ d \bm{\omega}_{i} \equiv \prod_{\kappa,\alpha} d \omega^{(\alpha)}_{i \kappa}$.
The integral of $\bm{w}_{i}$ can be evaluated by the saddle point method.
The saddle point equation is as follows:
\begin{equation}
\omega^{(\alpha)}_{i \kappa} = \varphi \sum_{\kappa,\alpha} W^{(\alpha)}_{i \kappa}.  \label{eq:A_saddle_w}
\end{equation}
Note that $\omega^{(\alpha)}_{i \kappa}$ does not depend on $\kappa$ and $\alpha$, consistent with the static approximation and assumption of the replica symmetric solution.
From Eqs. (\ref{eq:A_Li_sym}), (\ref{eq:A_Li_sym_1st}), (\ref{eq:A_Li_sym_2nd}), (\ref{eq:A_Li_sym_3rd}), and (\ref{eq:A_saddle_w}), we obtain
\begin{eqnarray}
\mathrm{Tr} e^{L_{i}} = \int d \bm{W}_{i} \exp &&\left[ \frac{\beta^{2}J^{2}Q^{2}}{2K^{2}} \left( \theta - 2 \varphi \right) \left( \sum_{\kappa,\alpha} W^{(\alpha)}_{i \kappa} \right)^{2} \right.  \nonumber  \\
&&+ \left. \frac{\beta^{2}J^{2}Q^{2}}{2K^{2}} \left( \varphi - \theta \right) \sum_{\alpha} \left( \sum_{\kappa} W^{(\alpha)}_{i \kappa} \right)^{2} + n \sum_{q} \int Du_{qi} \ln \int Dv_{qi} \mathrm{Tr} e^{L_{qi}} \right],
\end{eqnarray}
\end{widetext}
where
\begin{equation}
L_{qi} = \frac{\beta J_{\mathrm{T}}}{K} \sum_{\kappa} \sigma^{(\kappa)}_{qi} \sigma^{(\kappa+1)}_{qi} + \frac{\beta H_{qi}}{K} \sum_{\kappa} \sigma^{(\kappa)}_{qi},  \label{eq:A_Lqi}
\end{equation}
\begin{equation}
\frac{\beta J_{\mathrm{T}}}{K} = \frac{1}{2} \ln \coth \left( \frac{\beta \Gamma}{K} \right), 
\end{equation}
and
\begin{eqnarray}
H_{qi} &=& J \left[ \sqrt{ \xi_{q} - \theta } u_{qi} + \sqrt{ \left( \eta_{q} - \xi_{q} \right) - \left( \varphi - \theta \right) } v_{qi} \right]  \nonumber  \\
&+& \left( J_{0}m_{q} - \lambda M_{i} \right) + \frac{\beta J^{2}Q}{K} \varphi \left( \sum_{\kappa,\alpha} W^{(\alpha)}_{i \kappa} \right).
\end{eqnarray}
Here, we again apply the static approximation and assume the replica symmetric solution:
\begin{equation}
W_{i} \equiv W^{(\alpha)}_{i \kappa}.
\end{equation}
Neglecting terms that do not contribute in the limit of $n \to 0$ yields
\begin{eqnarray}
\mathrm{Tr} e^{L_{i}} &=& \int dW_{i} \exp \left\{ nQ \left[ \frac{\beta^{2}J^{2}Q}{2} \left( \varphi - \theta \right) W_{i}^{2} \right. \right.  \nonumber  \\
&+& \left. \left. \frac{1}{Q} \sum_{q} \int Du_{qi} \ln \int Dv_{qi} \mathrm{Tr} e^{L_{qi}} \right] \right\},  \label{eq:A_Tr_Li_n=0}
\end{eqnarray}
where
\begin{eqnarray}
H_{qi} &=& J \left[ \sqrt{ \xi_{q} - \theta } u_{qi} + \sqrt{ \left( \eta_{q} - \xi_{q} \right) - \left( \varphi - \theta \right) } v_{qi} \right]  \nonumber  \\
&+& \left( J_{0}m_{q} - \lambda M_{i} \right).  \label{eq:A_Hqi}
\end{eqnarray}
The integrals of $W_{i}$ in Eq. (\ref{eq:A_Tr_Li_n=0}) can be evaluated by the saddle point method, and we obtain
\begin{equation}
W_{i} = 0.  \label{eq:A_saddle_Wi}
\end{equation}
According to Eq. (\ref{eq:A_Lqi}), $\mathrm{Tr} e^{L_{qi}}$ is the partition function of the one-dimensional Ising model with the uniform interaction and magnetic field, which is given by the following equation:
\begin{equation}
\mathrm{Tr} e^{L_{qi}} = 2 e^{\beta J_{\mathrm{T}}} \cosh \beta \Xi_{qi},  \label{eq:A_1D_Ising}
\end{equation}
where
\begin{equation}
\Xi_{qi} = \sqrt{ H_{qi}^{2} + \Gamma^{2} },  \label{Appendix_C_free_energy2}
\end{equation}
in the limit of $K \to \infty$.
From Eqs. (\ref{eq:A_Tr_Li_n=0}), (\ref{eq:A_saddle_Wi}), and (\ref{eq:A_1D_Ising}), we obtain the free energy as follows:

\begin{widetext}
\begin{eqnarray}
- \beta f = - \frac{\beta J_{0}}{2Q} \sum_{q} m_{q}^{2} + \frac{\beta^{2}J^{2}}{4Q} \sum_{q} \xi_{q}^{2} &+& \frac{\beta^{2}J^{2}Q}{4} \theta^{2} - \frac{\beta^{2}J^{2}}{4Q} \sum_{q} \eta_{q}^{2} - \frac{\beta^{2}J^{2}Q}{4} \varphi^{2}  \nonumber  \\
&+& \frac{\beta \lambda}{2N} \sum_{i} M_{i}^{2} + \frac{1}{NQ} \sum_{q,i} \int Du_{qi} \ln \int Dv_{qi} 2 \cosh \beta \Xi_{qi}.  \label{Appendix_C_free_energy}
\end{eqnarray}
The order parameters are determined as the minimizer of free energy, and the saddle point equations are given by the following equation:
\begin{equation}
m_{q} = \frac{1}{N} \sum_{i} \int Du_{qi} \frac{ \displaystyle \int Dv_{qi} \frac{H_{qi}}{\Xi_{qi}} \sinh \beta \Xi_{qi} }{ \displaystyle \int Dv_{qi} \cosh \beta \Xi_{qi} },  \label{eq:glass_saddle_mq_sym}
\end{equation}
\begin{equation}
\xi_{q} = \frac{1}{N} \sum_{i} \int Du_{qi} \left( \frac{ \displaystyle \int Dv_{qi} \frac{H_{qi}}{\Xi_{qi}} \sinh \beta \Xi_{qi} }{ \displaystyle \int Dv_{qi} \cosh \beta \Xi_{qi} } \right)^{2},
\end{equation}
\begin{equation}
\eta_{q} = \frac{1}{N} \sum_{i} \int Du_{qi} \frac{ \displaystyle \int Dv_{qi} \left( \frac{H_{qi}^{2}}{\Xi_{qi}^{2}} \cosh \beta \Xi_{qi} + \frac{\Gamma^{2}}{\beta \Xi_{qi}^{3}} \sinh \beta \Xi_{qi} \right) }{ \displaystyle \int Dv_{qi} \cosh \beta \Xi_{qi} },
\end{equation}
\begin{equation}
M_{i} = \frac{1}{Q} \sum_{q} \int Du_{qi} \frac{ \displaystyle \int Dv_{qi} \frac{H_{qi}}{\Xi_{qi}} \sinh \beta \Xi_{qi} }{ \displaystyle \int Dv_{qi} \cosh \beta \Xi_{qi} }.  \label{eq:glass_saddle_Mi_sym}
\end{equation}
\begin{equation}
\theta = - \frac{1}{Q^{2}} \sum_{q} \xi_{q} \sim O \left( \frac{1}{Q} \right),
\end{equation}
\begin{equation}
\varphi = - \frac{1}{Q^{2}} \sum_{q} \eta_{q} \sim O \left( \frac{1}{Q} \right).
\end{equation}
Because $\theta$ and $\varphi$ are of order $Q^{-1}$, $\theta$ and $\varphi$ vanish in the limit of $Q \to \infty$.
By substituting $\theta = \varphi = 0$ into Eqs. (\ref{Appendix_C_free_energy}), (\ref{Appendix_C_free_energy2}), and (\ref{eq:A_Hqi}), we obtain Eqs. (\ref{eq:PG_free_energy}), (\ref{eq:glass_Xiqi}), and (\ref{eq:glass_Hqi_n=0}).
\end{widetext}

\bibliography{reference}

\providecommand{\noopsort}[1]{}\providecommand{\singleletter}[1]{#1}%
\begin{thebibliography}{49}%
\makeatletter
\providecommand \@ifxundefined [1]{%
 \@ifx{#1\undefined}
}%
\providecommand \@ifnum [1]{%
 \ifnum #1\expandafter \@firstoftwo
 \else \expandafter \@secondoftwo
 \fi
}%
\providecommand \@ifx [1]{%
 \ifx #1\expandafter \@firstoftwo
 \else \expandafter \@secondoftwo
 \fi
}%
\providecommand \natexlab [1]{#1}%
\providecommand \enquote  [1]{``#1''}%
\providecommand \bibnamefont  [1]{#1}%
\providecommand \bibfnamefont [1]{#1}%
\providecommand \citenamefont [1]{#1}%
\providecommand \href@noop [0]{\@secondoftwo}%
\providecommand \href [0]{\begingroup \@sanitize@url \@href}%
\providecommand \@href[1]{\@@startlink{#1}\@@href}%
\providecommand \@@href[1]{\endgroup#1\@@endlink}%
\providecommand \@sanitize@url [0]{\catcode `\\12\catcode `\$12\catcode
  `\&12\catcode `\#12\catcode `\^12\catcode `\_12\catcode `\%12\relax}%
\providecommand \@@startlink[1]{}%
\providecommand \@@endlink[0]{}%
\providecommand \url  [0]{\begingroup\@sanitize@url \@url }%
\providecommand \@url [1]{\endgroup\@href {#1}{\urlprefix }}%
\providecommand \urlprefix  [0]{URL }%
\providecommand \Eprint [0]{\href }%
\providecommand \doibase [0]{https://doi.org/}%
\providecommand \selectlanguage [0]{\@gobble}%
\providecommand \bibinfo  [0]{\@secondoftwo}%
\providecommand \bibfield  [0]{\@secondoftwo}%
\providecommand \translation [1]{[#1]}%
\providecommand \BibitemOpen [0]{}%
\providecommand \bibitemStop [0]{}%
\providecommand \bibitemNoStop [0]{.\EOS\space}%
\providecommand \EOS [0]{\spacefactor3000\relax}%
\providecommand \BibitemShut  [1]{\csname bibitem#1\endcsname}%
\let\auto@bib@innerbib\@empty
\bibitem [{\citenamefont {Lucas}(2014)}]{Ising_mapping}%
  \BibitemOpen
  \bibfield  {author} {\bibinfo {author} {\bibfnamefont {A.}~\bibnamefont
  {Lucas}},\ }\bibfield  {title} {\bibinfo {title} {Ising formulations of many
  np problems},\ }\href {https://doi.org/10.3389/fphy.2014.00005} {\bibfield
  {journal} {\bibinfo  {journal} {Front. Phys.}\ }\textbf {\bibinfo {volume}
  {2}},\ \bibinfo {pages} {5} (\bibinfo {year} {2014})}\BibitemShut {NoStop}%
\bibitem [{\citenamefont {Kirkpatrick}\ \emph {et~al.}(1983)\citenamefont
  {Kirkpatrick}, \citenamefont {Gelatt},\ and\ \citenamefont
  {Vecchi}}]{SA_original}%
  \BibitemOpen
  \bibfield  {author} {\bibinfo {author} {\bibfnamefont {S.}~\bibnamefont
  {Kirkpatrick}}, \bibinfo {author} {\bibfnamefont {C.~D.}\ \bibnamefont
  {Gelatt}},\ and\ \bibinfo {author} {\bibfnamefont {M.~P.}\ \bibnamefont
  {Vecchi}},\ }\bibfield  {title} {\bibinfo {title} {Optimization by simulated
  annealing},\ }\href {https://doi.org/10.1126/science.220.4598.671} {\bibfield
   {journal} {\bibinfo  {journal} {Science}\ }\textbf {\bibinfo {volume}
  {220}},\ \bibinfo {pages} {671} (\bibinfo {year} {1983})}\BibitemShut
  {NoStop}%
\bibitem [{\citenamefont {Kadowaki}\ and\ \citenamefont
  {Nishimori}(1998)}]{QA_original}%
  \BibitemOpen
  \bibfield  {author} {\bibinfo {author} {\bibfnamefont {T.}~\bibnamefont
  {Kadowaki}}\ and\ \bibinfo {author} {\bibfnamefont {H.}~\bibnamefont
  {Nishimori}},\ }\bibfield  {title} {\bibinfo {title} {Quantum annealing in
  the transverse ising model},\ }\href
  {https://doi.org/10.1103/PhysRevE.58.5355} {\bibfield  {journal} {\bibinfo
  {journal} {Phys. Rev. E}\ }\textbf {\bibinfo {volume} {58}},\ \bibinfo
  {pages} {5355} (\bibinfo {year} {1998})}\BibitemShut {NoStop}%
\bibitem [{\citenamefont {Farhi}\ \emph {et~al.}(2000)\citenamefont {Farhi},
  \citenamefont {Goldstone}, \citenamefont {Gutmann},\ and\ \citenamefont
  {Sipser}}]{AQC_original}%
  \BibitemOpen
  \bibfield  {author} {\bibinfo {author} {\bibfnamefont {E.}~\bibnamefont
  {Farhi}}, \bibinfo {author} {\bibfnamefont {J.}~\bibnamefont {Goldstone}},
  \bibinfo {author} {\bibfnamefont {S.}~\bibnamefont {Gutmann}},\ and\ \bibinfo
  {author} {\bibfnamefont {M.}~\bibnamefont {Sipser}},\ }\href@noop {}
  {}\bibinfo {howpublished} {e-print arXiv:quant-ph/0001106} (\bibinfo {year}
  {2000})\BibitemShut {NoStop}%
\bibitem [{\citenamefont {Morita}\ and\ \citenamefont
  {Nishimori}(2008)}]{adiabatic_theorem}%
  \BibitemOpen
  \bibfield  {author} {\bibinfo {author} {\bibfnamefont {S.}~\bibnamefont
  {Morita}}\ and\ \bibinfo {author} {\bibfnamefont {H.}~\bibnamefont
  {Nishimori}},\ }\bibfield  {title} {\bibinfo {title} {Mathematical foundation
  of quantum annealing},\ }\href {https://doi.org/10.1063/1.2995837} {\bibfield
   {journal} {\bibinfo  {journal} {J. Math. Phys.}\ }\textbf {\bibinfo {volume}
  {49}},\ \bibinfo {pages} {125210} (\bibinfo {year} {2008})}\BibitemShut
  {NoStop}%
\bibitem [{\citenamefont {Santoro}\ \emph {et~al.}(2002)\citenamefont
  {Santoro}, \citenamefont {Marto\v{n}\'{a}k}, \citenamefont {Tosatti},\ and\
  \citenamefont {Car}}]{QA_SA_compare1}%
  \BibitemOpen
  \bibfield  {author} {\bibinfo {author} {\bibfnamefont {G.~E.}\ \bibnamefont
  {Santoro}}, \bibinfo {author} {\bibfnamefont {R.}~\bibnamefont
  {Marto\v{n}\'{a}k}}, \bibinfo {author} {\bibfnamefont {E.}~\bibnamefont
  {Tosatti}},\ and\ \bibinfo {author} {\bibfnamefont {R.}~\bibnamefont {Car}},\
  }\bibfield  {title} {\bibinfo {title} {Theory of quantum annealing of an
  ising spin glass},\ }\href {https://doi.org/10.1126/science.1068774}
  {\bibfield  {journal} {\bibinfo  {journal} {Science}\ }\textbf {\bibinfo
  {volume} {295}},\ \bibinfo {pages} {2427} (\bibinfo {year}
  {2002})}\BibitemShut {NoStop}%
\bibitem [{\citenamefont {Marto\v{n}\'{a}k}\ \emph {et~al.}(2004)\citenamefont
  {Marto\v{n}\'{a}k}, \citenamefont {Santoro},\ and\ \citenamefont
  {Tosatti}}]{QA_SA_compare2}%
  \BibitemOpen
  \bibfield  {author} {\bibinfo {author} {\bibfnamefont {R.}~\bibnamefont
  {Marto\v{n}\'{a}k}}, \bibinfo {author} {\bibfnamefont {G.~E.}\ \bibnamefont
  {Santoro}},\ and\ \bibinfo {author} {\bibfnamefont {E.}~\bibnamefont
  {Tosatti}},\ }\bibfield  {title} {\bibinfo {title} {Quantum annealing of
  traveling-salesman problem},\ }\href
  {https://doi.org/10.1103/PhysRevE.70.057701} {\bibfield  {journal} {\bibinfo
  {journal} {Phys. Rev. E}\ }\textbf {\bibinfo {volume} {70}},\ \bibinfo
  {pages} {057701} (\bibinfo {year} {2004})}\BibitemShut {NoStop}%
\bibitem [{\citenamefont {Stella}\ \emph {et~al.}(2005)\citenamefont {Stella},
  \citenamefont {Santoro},\ and\ \citenamefont {Tosatti}}]{QA_SA_compare3}%
  \BibitemOpen
  \bibfield  {author} {\bibinfo {author} {\bibfnamefont {L.}~\bibnamefont
  {Stella}}, \bibinfo {author} {\bibfnamefont {G.~E.}\ \bibnamefont
  {Santoro}},\ and\ \bibinfo {author} {\bibfnamefont {E.}~\bibnamefont
  {Tosatti}},\ }\bibfield  {title} {\bibinfo {title} {Optimization by quantum
  annealing: Lessons from simple cases},\ }\href
  {https://doi.org/10.1103/PhysRevB.72.014303} {\bibfield  {journal} {\bibinfo
  {journal} {Phys. Rev. B}\ }\textbf {\bibinfo {volume} {72}},\ \bibinfo
  {pages} {014303} (\bibinfo {year} {2005})}\BibitemShut {NoStop}%
\bibitem [{\citenamefont {Battaglia}\ \emph {et~al.}(2005)\citenamefont
  {Battaglia}, \citenamefont {Santoro},\ and\ \citenamefont
  {Tosatti}}]{QA_SA_compare4}%
  \BibitemOpen
  \bibfield  {author} {\bibinfo {author} {\bibfnamefont {D.~A.}\ \bibnamefont
  {Battaglia}}, \bibinfo {author} {\bibfnamefont {G.~E.}\ \bibnamefont
  {Santoro}},\ and\ \bibinfo {author} {\bibfnamefont {E.}~\bibnamefont
  {Tosatti}},\ }\bibfield  {title} {\bibinfo {title} {Optimization by quantum
  annealing: Lessons from hard satisfiability problems},\ }\href
  {https://doi.org/10.1103/PhysRevE.71.066707} {\bibfield  {journal} {\bibinfo
  {journal} {Phys. Rev. E}\ }\textbf {\bibinfo {volume} {71}},\ \bibinfo
  {pages} {066707} (\bibinfo {year} {2005})}\BibitemShut {NoStop}%
\bibitem [{\citenamefont {Zanca}\ and\ \citenamefont
  {Santoro}(2016)}]{QA_SA_compare5}%
  \BibitemOpen
  \bibfield  {author} {\bibinfo {author} {\bibfnamefont {T.}~\bibnamefont
  {Zanca}}\ and\ \bibinfo {author} {\bibfnamefont {G.~E.}\ \bibnamefont
  {Santoro}},\ }\bibfield  {title} {\bibinfo {title} {Quantum annealing speedup
  over simulated annealing on random ising chains},\ }\href
  {https://doi.org/10.1103/PhysRevB.93.224431} {\bibfield  {journal} {\bibinfo
  {journal} {Phys. Rev. B}\ }\textbf {\bibinfo {volume} {93}},\ \bibinfo
  {pages} {224431} (\bibinfo {year} {2016})}\BibitemShut {NoStop}%
\bibitem [{\citenamefont {Wauters}\ \emph {et~al.}(2017)\citenamefont
  {Wauters}, \citenamefont {Fazio}, \citenamefont {Nishimori},\ and\
  \citenamefont {Santoro}}]{QA_SA_compare6}%
  \BibitemOpen
  \bibfield  {author} {\bibinfo {author} {\bibfnamefont {M.~M.}\ \bibnamefont
  {Wauters}}, \bibinfo {author} {\bibfnamefont {R.}~\bibnamefont {Fazio}},
  \bibinfo {author} {\bibfnamefont {H.}~\bibnamefont {Nishimori}},\ and\
  \bibinfo {author} {\bibfnamefont {G.~E.}\ \bibnamefont {Santoro}},\
  }\bibfield  {title} {\bibinfo {title} {Direct comparison of quantum and
  simulated annealing on a fully connected ising ferromagnet},\ }\href
  {https://doi.org/10.1103/PhysRevA.96.022326} {\bibfield  {journal} {\bibinfo
  {journal} {Phys. Rev. A}\ }\textbf {\bibinfo {volume} {96}},\ \bibinfo
  {pages} {022326} (\bibinfo {year} {2017})}\BibitemShut {NoStop}%
\bibitem [{\citenamefont {Ohzeki}(2017)}]{non-stoquastic_QMC}%
  \BibitemOpen
  \bibfield  {author} {\bibinfo {author} {\bibfnamefont {M.}~\bibnamefont
  {Ohzeki}},\ }\bibfield  {title} {\bibinfo {title} {Quantum monte carlo
  simulation of a particular class of non-stoquastic hamiltonians in quantum
  annealing},\ }\href {https://doi.org/10.1038/srep41186} {\bibfield  {journal}
  {\bibinfo  {journal} {Sci. Rep.}\ }\textbf {\bibinfo {volume} {7}},\ \bibinfo
  {pages} {41186} (\bibinfo {year} {2017})}\BibitemShut {NoStop}%
\bibitem [{\citenamefont {Seki}\ and\ \citenamefont
  {Nishimori}(2012)}]{XX_p-spin}%
  \BibitemOpen
  \bibfield  {author} {\bibinfo {author} {\bibfnamefont {Y.}~\bibnamefont
  {Seki}}\ and\ \bibinfo {author} {\bibfnamefont {H.}~\bibnamefont
  {Nishimori}},\ }\bibfield  {title} {\bibinfo {title} {Quantum annealing with
  antiferromagnetic fluctuations},\ }\href
  {https://doi.org/10.1103/PhysRevE.85.051112} {\bibfield  {journal} {\bibinfo
  {journal} {Phys. Rev. E}\ }\textbf {\bibinfo {volume} {85}},\ \bibinfo
  {pages} {051112} (\bibinfo {year} {2012})}\BibitemShut {NoStop}%
\bibitem [{\citenamefont {Seki}\ and\ \citenamefont
  {Nishimori}(2015)}]{XX_Hopfield}%
  \BibitemOpen
  \bibfield  {author} {\bibinfo {author} {\bibfnamefont {Y.}~\bibnamefont
  {Seki}}\ and\ \bibinfo {author} {\bibfnamefont {H.}~\bibnamefont
  {Nishimori}},\ }\bibfield  {title} {\bibinfo {title} {Quantum annealing with
  antiferromagnetic transverse interactions for the hopfield model},\ }\href
  {https://doi.org/10.1088/1751-8113/48/33/335301} {\bibfield  {journal}
  {\bibinfo  {journal} {J. Phys. A}\ }\textbf {\bibinfo {volume} {48}},\
  \bibinfo {pages} {335301} (\bibinfo {year} {2015})}\BibitemShut {NoStop}%
\bibitem [{\citenamefont {Ohkuwa}\ and\ \citenamefont
  {Nishimori}(2017)}]{XX_p-spin_gap}%
  \BibitemOpen
  \bibfield  {author} {\bibinfo {author} {\bibfnamefont {M.}~\bibnamefont
  {Ohkuwa}}\ and\ \bibinfo {author} {\bibfnamefont {H.}~\bibnamefont
  {Nishimori}},\ }\bibfield  {title} {\bibinfo {title} {Exact expression of the
  energy gap at first-order phase transitions of the fully connected p-body
  transverse-field ising model with transverse interactions},\ }\href
  {https://doi.org/10.7566/JPSJ.86.114004} {\bibfield  {journal} {\bibinfo
  {journal} {J. Phys. Soc. Jpn.}\ }\textbf {\bibinfo {volume} {86}},\ \bibinfo
  {pages} {114004} (\bibinfo {year} {2017})}\BibitemShut {NoStop}%
\bibitem [{\citenamefont {Arai}\ \emph {et~al.}(2018)\citenamefont {Arai},
  \citenamefont {Ohzeki},\ and\ \citenamefont {Tanaka}}]{XX_QMC}%
  \BibitemOpen
  \bibfield  {author} {\bibinfo {author} {\bibfnamefont {S.}~\bibnamefont
  {Arai}}, \bibinfo {author} {\bibfnamefont {M.}~\bibnamefont {Ohzeki}},\ and\
  \bibinfo {author} {\bibfnamefont {K.}~\bibnamefont {Tanaka}},\ }\href@noop {}
  {}\bibinfo {howpublished} {arXiv:1810.09943} (\bibinfo {year}
  {2018})\BibitemShut {NoStop}%
\bibitem [{\citenamefont {Okada}\ \emph {et~al.}(2019)\citenamefont {Okada},
  \citenamefont {Ohzeki},\ and\ \citenamefont {Tanaka}}]{XX_one-dimension}%
  \BibitemOpen
  \bibfield  {author} {\bibinfo {author} {\bibfnamefont {S.}~\bibnamefont
  {Okada}}, \bibinfo {author} {\bibfnamefont {M.}~\bibnamefont {Ohzeki}},\ and\
  \bibinfo {author} {\bibfnamefont {K.}~\bibnamefont {Tanaka}},\ }\bibfield
  {title} {\bibinfo {title} {Phase diagrams of one-dimensional ising and xy
  models with fully connected ferromagnetic and anti-ferromagnetic quantum
  fluctuations},\ }\href {https://doi.org/10.7566/JPSJ.88.024802} {\bibfield
  {journal} {\bibinfo  {journal} {J. Phys. Soc. Jpn.}\ }\textbf {\bibinfo
  {volume} {88}},\ \bibinfo {pages} {024802} (\bibinfo {year}
  {2019})}\BibitemShut {NoStop}%
\bibitem [{\citenamefont {Johnson}\ \emph {et~al.}(2011)\citenamefont {Johnson}
  \emph {et~al.}}]{D-Wave_machine}%
  \BibitemOpen
  \bibfield  {author} {\bibinfo {author} {\bibfnamefont {M.~W.}\ \bibnamefont
  {Johnson}} \emph {et~al.},\ }\bibfield  {title} {\bibinfo {title} {Quantum
  annealing with manufactured spins},\ }\href
  {https://doi.org/10.1038/nature10012} {\bibfield  {journal} {\bibinfo
  {journal} {Nature}\ }\textbf {\bibinfo {volume} {473}},\ \bibinfo {pages}
  {194} (\bibinfo {year} {2011})}\BibitemShut {NoStop}%
\bibitem [{\citenamefont {R{\o}nnow}\ \emph {et~al.}(2014)\citenamefont
  {R{\o}nnow} \emph {et~al.}}]{D-Wave_compare1}%
  \BibitemOpen
  \bibfield  {author} {\bibinfo {author} {\bibfnamefont {T.~F.}\ \bibnamefont
  {R{\o}nnow}} \emph {et~al.},\ }\bibfield  {title} {\bibinfo {title} {Defining
  and detecting quantum speedup},\ }\href
  {https://doi.org/10.1126/science.1252319} {\bibfield  {journal} {\bibinfo
  {journal} {science}\ }\textbf {\bibinfo {volume} {345}},\ \bibinfo {pages}
  {420} (\bibinfo {year} {2014})}\BibitemShut {NoStop}%
\bibitem [{\citenamefont {Katzgraber}\ \emph {et~al.}(2015)\citenamefont
  {Katzgraber}, \citenamefont {Hamze}, \citenamefont {Zhu}, \citenamefont
  {Ochoa},\ and\ \citenamefont {Munoz-Bauza}}]{D-Wave_compare2}%
  \BibitemOpen
  \bibfield  {author} {\bibinfo {author} {\bibfnamefont {H.~G.}\ \bibnamefont
  {Katzgraber}}, \bibinfo {author} {\bibfnamefont {F.}~\bibnamefont {Hamze}},
  \bibinfo {author} {\bibfnamefont {Z.}~\bibnamefont {Zhu}}, \bibinfo {author}
  {\bibfnamefont {A.~J.}\ \bibnamefont {Ochoa}},\ and\ \bibinfo {author}
  {\bibfnamefont {H.}~\bibnamefont {Munoz-Bauza}},\ }\bibfield  {title}
  {\bibinfo {title} {Seeking quantum speedup through spin glasses: The good,
  the bad, and the ugly},\ }\href {https://doi.org/10.1103/PhysRevX.5.031026}
  {\bibfield  {journal} {\bibinfo  {journal} {Phys. Rev. X}\ }\textbf {\bibinfo
  {volume} {5}},\ \bibinfo {pages} {031026} (\bibinfo {year}
  {2015})}\BibitemShut {NoStop}%
\bibitem [{\citenamefont {Denchev}\ \emph {et~al.}(2016)\citenamefont
  {Denchev}, \citenamefont {Boixo}, \citenamefont {Isakov}, \citenamefont
  {Ding}, \citenamefont {Babbush}, \citenamefont {Smelyanskiy}, \citenamefont
  {Martinis},\ and\ \citenamefont {Neven}}]{D-Wave_compare3}%
  \BibitemOpen
  \bibfield  {author} {\bibinfo {author} {\bibfnamefont {V.~S.}\ \bibnamefont
  {Denchev}}, \bibinfo {author} {\bibfnamefont {S.}~\bibnamefont {Boixo}},
  \bibinfo {author} {\bibfnamefont {S.~V.}\ \bibnamefont {Isakov}}, \bibinfo
  {author} {\bibfnamefont {N.}~\bibnamefont {Ding}}, \bibinfo {author}
  {\bibfnamefont {R.}~\bibnamefont {Babbush}}, \bibinfo {author} {\bibfnamefont
  {V.}~\bibnamefont {Smelyanskiy}}, \bibinfo {author} {\bibfnamefont
  {J.}~\bibnamefont {Martinis}},\ and\ \bibinfo {author} {\bibfnamefont
  {H.}~\bibnamefont {Neven}},\ }\bibfield  {title} {\bibinfo {title} {What is
  the computational value of finite range tunneling?},\ }\href
  {https://doi.org/10.1103/PhysRevX.6.031015} {\bibfield  {journal} {\bibinfo
  {journal} {Phys. Rev. X}\ }\textbf {\bibinfo {volume} {6}},\ \bibinfo {pages}
  {031015} (\bibinfo {year} {2016})}\BibitemShut {NoStop}%
\bibitem [{\citenamefont {Wang}\ \emph {et~al.}(2016)\citenamefont {Wang},
  \citenamefont {Chen},\ and\ \citenamefont
  {Jonckheere}}]{D-Wave_application1}%
  \BibitemOpen
  \bibfield  {author} {\bibinfo {author} {\bibfnamefont {C.}~\bibnamefont
  {Wang}}, \bibinfo {author} {\bibfnamefont {H.}~\bibnamefont {Chen}},\ and\
  \bibinfo {author} {\bibfnamefont {E.}~\bibnamefont {Jonckheere}},\ }\bibfield
   {title} {\bibinfo {title} {Quantum versus simulated annealing in wireless
  interference network optimization},\ }\href
  {https://doi.org/10.1038/srep25797} {\bibfield  {journal} {\bibinfo
  {journal} {Sci. Rep.}\ }\textbf {\bibinfo {volume} {6}},\ \bibinfo {pages}
  {25797} (\bibinfo {year} {2016})}\BibitemShut {NoStop}%
\bibitem [{\citenamefont {Rosenberg}\ \emph {et~al.}(2016)\citenamefont
  {Rosenberg}, \citenamefont {Haghnegahdar}, \citenamefont {Goddard},
  \citenamefont {Carr}, \citenamefont {Wu},\ and\ \citenamefont
  {de~Prado}}]{D-Wave_application2}%
  \BibitemOpen
  \bibfield  {author} {\bibinfo {author} {\bibfnamefont {G.}~\bibnamefont
  {Rosenberg}}, \bibinfo {author} {\bibfnamefont {P.}~\bibnamefont
  {Haghnegahdar}}, \bibinfo {author} {\bibfnamefont {P.}~\bibnamefont
  {Goddard}}, \bibinfo {author} {\bibfnamefont {P.}~\bibnamefont {Carr}},
  \bibinfo {author} {\bibfnamefont {K.}~\bibnamefont {Wu}},\ and\ \bibinfo
  {author} {\bibfnamefont {M.~L.}\ \bibnamefont {de~Prado}},\ }\bibfield
  {title} {\bibinfo {title} {Solving the optimal trading trajectory problem
  using a quantum annealer},\ }\href
  {https://doi.org/10.1109/JSTSP.2016.2574703} {\bibfield  {journal} {\bibinfo
  {journal} {IEEE Journal of Selected Topics in Signal Processing}\ }\textbf
  {\bibinfo {volume} {10}},\ \bibinfo {pages} {1053} (\bibinfo {year}
  {2016})}\BibitemShut {NoStop}%
\bibitem [{\citenamefont {Boyda}\ \emph {et~al.}(2017)\citenamefont {Boyda},
  \citenamefont {Basu}, \citenamefont {Ganguly}, \citenamefont {Michaelis},
  \citenamefont {Mukhopadhyay},\ and\ \citenamefont
  {Nemani}}]{D-Wave_application3}%
  \BibitemOpen
  \bibfield  {author} {\bibinfo {author} {\bibfnamefont {E.}~\bibnamefont
  {Boyda}}, \bibinfo {author} {\bibfnamefont {S.}~\bibnamefont {Basu}},
  \bibinfo {author} {\bibfnamefont {S.}~\bibnamefont {Ganguly}}, \bibinfo
  {author} {\bibfnamefont {A.}~\bibnamefont {Michaelis}}, \bibinfo {author}
  {\bibfnamefont {S.}~\bibnamefont {Mukhopadhyay}},\ and\ \bibinfo {author}
  {\bibfnamefont {R.~R.}\ \bibnamefont {Nemani}},\ }\bibfield  {title}
  {\bibinfo {title} {Deploying a quantum annealing processor to detect tree
  cover in aerial imagery of california},\ }\href
  {https://doi.org/10.1371/journal.pone.0172505} {\bibfield  {journal}
  {\bibinfo  {journal} {PLoS ONE}\ ,\ \bibinfo {pages} {12(2): e0172505}}
  (\bibinfo {year} {2017})}\BibitemShut {NoStop}%
\bibitem [{\citenamefont {O'Malley}\ \emph {et~al.}(2017)\citenamefont
  {O'Malley}, \citenamefont {Vesselinov}, \citenamefont {Alexandrov},\ and\
  \citenamefont {Alexandrov}}]{D-Wave_application4}%
  \BibitemOpen
  \bibfield  {author} {\bibinfo {author} {\bibfnamefont {D.}~\bibnamefont
  {O'Malley}}, \bibinfo {author} {\bibfnamefont {V.~V.}\ \bibnamefont
  {Vesselinov}}, \bibinfo {author} {\bibfnamefont {B.~S.}\ \bibnamefont
  {Alexandrov}},\ and\ \bibinfo {author} {\bibfnamefont {L.~B.}\ \bibnamefont
  {Alexandrov}},\ }\href@noop {} {}\bibinfo {howpublished} {e-print
  arXiv:1704.01605} (\bibinfo {year} {2017})\BibitemShut {NoStop}%
\bibitem [{\citenamefont {Neukart}\ \emph {et~al.}(2017)\citenamefont
  {Neukart}, \citenamefont {Compostella}, \citenamefont {Seidel}, \citenamefont
  {Dollen}, \citenamefont {Yarkoni},\ and\ \citenamefont
  {Parney}}]{D-Wave_application5}%
  \BibitemOpen
  \bibfield  {author} {\bibinfo {author} {\bibfnamefont {F.}~\bibnamefont
  {Neukart}}, \bibinfo {author} {\bibfnamefont {G.}~\bibnamefont
  {Compostella}}, \bibinfo {author} {\bibfnamefont {C.}~\bibnamefont {Seidel}},
  \bibinfo {author} {\bibfnamefont {D.~V.}\ \bibnamefont {Dollen}}, \bibinfo
  {author} {\bibfnamefont {S.}~\bibnamefont {Yarkoni}},\ and\ \bibinfo {author}
  {\bibfnamefont {B.}~\bibnamefont {Parney}},\ }\bibfield  {title} {\bibinfo
  {title} {Traffic flow optimization using a quantum annealer},\ }\href
  {https://doi.org/10.3389/fict.2017.00029} {\bibfield  {journal} {\bibinfo
  {journal} {Frontiers in ICT}\ }\textbf {\bibinfo {volume} {4}},\ \bibinfo
  {pages} {29} (\bibinfo {year} {2017})}\BibitemShut {NoStop}%
\bibitem [{\citenamefont {Baldassi}\ and\ \citenamefont
  {Zecchina}(2018)}]{D-Wave_application6}%
  \BibitemOpen
  \bibfield  {author} {\bibinfo {author} {\bibfnamefont {C.}~\bibnamefont
  {Baldassi}}\ and\ \bibinfo {author} {\bibfnamefont {R.}~\bibnamefont
  {Zecchina}},\ }\bibfield  {title} {\bibinfo {title} {Efficiency of quantum
  vs. classical annealing in nonconvex learning problems},\ }\href
  {https://doi.org/10.1073/pnas.1711456115} {\bibfield  {journal} {\bibinfo
  {journal} {Proceedings of the National Academy of Sciences}\ }\textbf
  {\bibinfo {volume} {115}},\ \bibinfo {pages} {1457} (\bibinfo {year}
  {2018})}\BibitemShut {NoStop}%
\bibitem [{\citenamefont {Yarkoni}\ \emph {et~al.}(2018)\citenamefont
  {Yarkoni}, \citenamefont {Plaat},\ and\ \citenamefont
  {Back}}]{D-Wave_application7}%
  \BibitemOpen
  \bibfield  {author} {\bibinfo {author} {\bibfnamefont {S.}~\bibnamefont
  {Yarkoni}}, \bibinfo {author} {\bibfnamefont {A.}~\bibnamefont {Plaat}},\
  and\ \bibinfo {author} {\bibfnamefont {T.}~\bibnamefont {Back}},\ }\bibfield
  {title} {\bibinfo {title} {First results solving arbitrarily structured
  maximum independent set problems using quantum annealing},\ }in\ \href
  {https://doi.org/10.1109/CEC.2018.8477865} {\emph {\bibinfo {booktitle} {2018
  IEEE Congress on Evolutionary Computation (CEC)}}}\ (\bibinfo {year} {2018})\
  pp.\ \bibinfo {pages} {1--6}\BibitemShut {NoStop}%
\bibitem [{\citenamefont {Adachi}\ and\ \citenamefont
  {Henderson}(2015)}]{D-Wave_application8}%
  \BibitemOpen
  \bibfield  {author} {\bibinfo {author} {\bibfnamefont {S.~H.}\ \bibnamefont
  {Adachi}}\ and\ \bibinfo {author} {\bibfnamefont {M.~P.}\ \bibnamefont
  {Henderson}},\ }\href@noop {} {}\bibinfo {howpublished} {e-print
  arXiv:1510.06356} (\bibinfo {year} {2015})\BibitemShut {NoStop}%
\bibitem [{\citenamefont {Amin}\ \emph {et~al.}(2018)\citenamefont {Amin},
  \citenamefont {Andriyash}, \citenamefont {Rolfe}, \citenamefont
  {Kulchytskyy},\ and\ \citenamefont {Melko}}]{D-Wave_application9}%
  \BibitemOpen
  \bibfield  {author} {\bibinfo {author} {\bibfnamefont {M.~H.}\ \bibnamefont
  {Amin}}, \bibinfo {author} {\bibfnamefont {E.}~\bibnamefont {Andriyash}},
  \bibinfo {author} {\bibfnamefont {J.}~\bibnamefont {Rolfe}}, \bibinfo
  {author} {\bibfnamefont {B.}~\bibnamefont {Kulchytskyy}},\ and\ \bibinfo
  {author} {\bibfnamefont {R.}~\bibnamefont {Melko}},\ }\bibfield  {title}
  {\bibinfo {title} {Quantum boltzmann machine},\ }\href
  {https://doi.org/10.1103/PhysRevX.8.021050} {\bibfield  {journal} {\bibinfo
  {journal} {Phys. Rev. X}\ }\textbf {\bibinfo {volume} {8}},\ \bibinfo {pages}
  {021050} (\bibinfo {year} {2018})}\BibitemShut {NoStop}%
\bibitem [{\citenamefont {Benedetti}\ \emph {et~al.}(2017)\citenamefont
  {Benedetti}, \citenamefont {Realpe-G\'{o}mez}, \citenamefont {Biswas},\ and\
  \citenamefont {Perdomo-Ortiz}}]{D-Wave_application10}%
  \BibitemOpen
  \bibfield  {author} {\bibinfo {author} {\bibfnamefont {M.}~\bibnamefont
  {Benedetti}}, \bibinfo {author} {\bibfnamefont {J.}~\bibnamefont
  {Realpe-G\'{o}mez}}, \bibinfo {author} {\bibfnamefont {R.}~\bibnamefont
  {Biswas}},\ and\ \bibinfo {author} {\bibfnamefont {A.}~\bibnamefont
  {Perdomo-Ortiz}},\ }\bibfield  {title} {\bibinfo {title} {Quantum-assisted
  learning of hardware-embedded probabilistic graphical models},\ }\href
  {https://doi.org/10.1103/PhysRevX.7.041052} {\bibfield  {journal} {\bibinfo
  {journal} {Phys. Rev. X}\ }\textbf {\bibinfo {volume} {7}},\ \bibinfo {pages}
  {041052} (\bibinfo {year} {2017})}\BibitemShut {NoStop}%
\bibitem [{\citenamefont {Harris}\ \emph {et~al.}(2018)\citenamefont {Harris}
  \emph {et~al.}}]{D-Wave_application11}%
  \BibitemOpen
  \bibfield  {author} {\bibinfo {author} {\bibfnamefont {R.}~\bibnamefont
  {Harris}} \emph {et~al.},\ }\bibfield  {title} {\bibinfo {title} {Phase
  transitions in a programmable quantum spin glass simulator},\ }\href
  {https://doi.org/10.1126/science.aat2025} {\bibfield  {journal} {\bibinfo
  {journal} {Science}\ }\textbf {\bibinfo {volume} {361}},\ \bibinfo {pages}
  {162} (\bibinfo {year} {2018})}\BibitemShut {NoStop}%
\bibitem [{\citenamefont {King}\ \emph {et~al.}(2018)\citenamefont {King} \emph
  {et~al.}}]{D-Wave_application12}%
  \BibitemOpen
  \bibfield  {author} {\bibinfo {author} {\bibfnamefont {A.~D.}\ \bibnamefont
  {King}} \emph {et~al.},\ }\bibfield  {title} {\bibinfo {title} {Observation
  of topological phenomena in a programmable lattice of 1,800 qubits},\ }\href
  {https://doi.org/10.1038/s41586-018-0410-x} {\bibfield  {journal} {\bibinfo
  {journal} {Nature}\ }\textbf {\bibinfo {volume} {560}},\ \bibinfo {pages}
  {456} (\bibinfo {year} {2018})}\BibitemShut {NoStop}%
\bibitem [{\citenamefont {Streif}\ \emph {et~al.}(2018)\citenamefont {Streif},
  \citenamefont {Neukart},\ and\ \citenamefont {Leib}}]{D-Wave_application13}%
  \BibitemOpen
  \bibfield  {author} {\bibinfo {author} {\bibfnamefont {M.}~\bibnamefont
  {Streif}}, \bibinfo {author} {\bibfnamefont {F.}~\bibnamefont {Neukart}},\
  and\ \bibinfo {author} {\bibfnamefont {M.}~\bibnamefont {Leib}},\ }\href@noop
  {} {}\bibinfo {howpublished} {e-print arXiv:1811.05256} (\bibinfo {year}
  {2018})\BibitemShut {NoStop}%
\bibitem [{\citenamefont {Ohzeki}\ \emph {et~al.}(2018)\citenamefont {Ohzeki},
  \citenamefont {Miki}, \citenamefont {Miyama},\ and\ \citenamefont
  {Terabe}}]{D-Wave_application14}%
  \BibitemOpen
  \bibfield  {author} {\bibinfo {author} {\bibfnamefont {M.}~\bibnamefont
  {Ohzeki}}, \bibinfo {author} {\bibfnamefont {A.}~\bibnamefont {Miki}},
  \bibinfo {author} {\bibfnamefont {M.~J.}\ \bibnamefont {Miyama}},\ and\
  \bibinfo {author} {\bibfnamefont {M.}~\bibnamefont {Terabe}},\ }\href@noop {}
  {}\bibinfo {howpublished} {arXiv:1812.01532} (\bibinfo {year}
  {2018})\BibitemShut {NoStop}%
\bibitem [{\citenamefont {Kitai}\ \emph {et~al.}(2019)\citenamefont {Kitai},
  \citenamefont {Guo}, \citenamefont {Ju}, \citenamefont {Tanaka},
  \citenamefont {Tsuda}, \citenamefont {Shiomi},\ and\ \citenamefont
  {Tamura}}]{D-Wave_application15}%
  \BibitemOpen
  \bibfield  {author} {\bibinfo {author} {\bibfnamefont {K.}~\bibnamefont
  {Kitai}}, \bibinfo {author} {\bibfnamefont {J.}~\bibnamefont {Guo}}, \bibinfo
  {author} {\bibfnamefont {S.}~\bibnamefont {Ju}}, \bibinfo {author}
  {\bibfnamefont {S.}~\bibnamefont {Tanaka}}, \bibinfo {author} {\bibfnamefont
  {K.}~\bibnamefont {Tsuda}}, \bibinfo {author} {\bibfnamefont
  {J.}~\bibnamefont {Shiomi}},\ and\ \bibinfo {author} {\bibfnamefont
  {R.}~\bibnamefont {Tamura}},\ }\href@noop {} {}\bibinfo {howpublished}
  {arXiv:1902.06573} (\bibinfo {year} {2019})\BibitemShut {NoStop}%
\bibitem [{\citenamefont {Irie}\ \emph {et~al.}(2019)\citenamefont {Irie},
  \citenamefont {Wongpaisarnsin}, \citenamefont {Terabe}, \citenamefont
  {Miki},\ and\ \citenamefont {Taguchi}}]{D-Wave_application16}%
  \BibitemOpen
  \bibfield  {author} {\bibinfo {author} {\bibfnamefont {H.}~\bibnamefont
  {Irie}}, \bibinfo {author} {\bibfnamefont {G.}~\bibnamefont
  {Wongpaisarnsin}}, \bibinfo {author} {\bibfnamefont {M.}~\bibnamefont
  {Terabe}}, \bibinfo {author} {\bibfnamefont {A.}~\bibnamefont {Miki}},\ and\
  \bibinfo {author} {\bibfnamefont {S.}~\bibnamefont {Taguchi}},\ }\bibfield
  {title} {\bibinfo {title} {Quantum annealing of vehicle routing problem with
  time, state and capacity},\ }in\ \href
  {https://doi.org/10.1007/978-3-030-14082-3_13} {\emph {\bibinfo {booktitle}
  {Quantum Technology and Optimization Problems}}}\ (\bibinfo {year} {2019})\
  pp.\ \bibinfo {pages} {145--156}\BibitemShut {NoStop}%
\bibitem [{\citenamefont {Wu}(1982)}]{Potts}%
  \BibitemOpen
  \bibfield  {author} {\bibinfo {author} {\bibfnamefont {F.~Y.}\ \bibnamefont
  {Wu}},\ }\bibfield  {title} {\bibinfo {title} {The potts model},\ }\href
  {https://doi.org/10.1103/RevModPhys.54.235} {\bibfield  {journal} {\bibinfo
  {journal} {Rev. Mod. Phys.}\ }\textbf {\bibinfo {volume} {54}},\ \bibinfo
  {pages} {235} (\bibinfo {year} {1982})}\BibitemShut {NoStop}%
\bibitem [{\citenamefont {Aramon}\ \emph {et~al.}(2018)\citenamefont {Aramon},
  \citenamefont {Rosenberg}, \citenamefont {Valiante}, \citenamefont
  {Miyazawa}, \citenamefont {Tamura},\ and\ \citenamefont
  {Katzgraber}}]{digital_annealer}%
  \BibitemOpen
  \bibfield  {author} {\bibinfo {author} {\bibfnamefont {M.}~\bibnamefont
  {Aramon}}, \bibinfo {author} {\bibfnamefont {G.}~\bibnamefont {Rosenberg}},
  \bibinfo {author} {\bibfnamefont {E.}~\bibnamefont {Valiante}}, \bibinfo
  {author} {\bibfnamefont {T.}~\bibnamefont {Miyazawa}}, \bibinfo {author}
  {\bibfnamefont {H.}~\bibnamefont {Tamura}},\ and\ \bibinfo {author}
  {\bibfnamefont {H.~G.}\ \bibnamefont {Katzgraber}},\ }\href@noop {}
  {}\bibinfo {howpublished} {arXiv:1806.08815} (\bibinfo {year}
  {2018})\BibitemShut {NoStop}%
\bibitem [{\citenamefont {J\"org}\ \emph {et~al.}(2008)\citenamefont {J\"org},
  \citenamefont {Krzakala}, \citenamefont {Kurchan},\ and\ \citenamefont
  {Maggs}}]{gap_first-order1}%
  \BibitemOpen
  \bibfield  {author} {\bibinfo {author} {\bibfnamefont {T.}~\bibnamefont
  {J\"org}}, \bibinfo {author} {\bibfnamefont {F.}~\bibnamefont {Krzakala}},
  \bibinfo {author} {\bibfnamefont {J.}~\bibnamefont {Kurchan}},\ and\ \bibinfo
  {author} {\bibfnamefont {A.~C.}\ \bibnamefont {Maggs}},\ }\bibfield  {title}
  {\bibinfo {title} {Simple glass models and their quantum annealing},\ }\href
  {https://doi.org/10.1103/PhysRevLett.101.147204} {\bibfield  {journal}
  {\bibinfo  {journal} {Phys. Rev. Lett.}\ }\textbf {\bibinfo {volume} {101}},\
  \bibinfo {pages} {147204} (\bibinfo {year} {2008})}\BibitemShut {NoStop}%
\bibitem [{\citenamefont {J\"org}\ \emph
  {et~al.}(2010{\natexlab{a}})\citenamefont {J\"org}, \citenamefont {Krzakala},
  \citenamefont {Semerjian},\ and\ \citenamefont {Zamponi}}]{gap_first-order2}%
  \BibitemOpen
  \bibfield  {author} {\bibinfo {author} {\bibfnamefont {T.}~\bibnamefont
  {J\"org}}, \bibinfo {author} {\bibfnamefont {F.}~\bibnamefont {Krzakala}},
  \bibinfo {author} {\bibfnamefont {G.}~\bibnamefont {Semerjian}},\ and\
  \bibinfo {author} {\bibfnamefont {F.}~\bibnamefont {Zamponi}},\ }\bibfield
  {title} {\bibinfo {title} {First-order transitions and the performance of
  quantum algorithms in random optimization problems},\ }\href
  {https://doi.org/10.1103/PhysRevLett.104.207206} {\bibfield  {journal}
  {\bibinfo  {journal} {Phys. Rev. Lett.}\ }\textbf {\bibinfo {volume} {104}},\
  \bibinfo {pages} {207206} (\bibinfo {year} {2010}{\natexlab{a}})}\BibitemShut
  {NoStop}%
\bibitem [{\citenamefont {J\"org}\ \emph
  {et~al.}(2010{\natexlab{b}})\citenamefont {J\"org}, \citenamefont {Krzakala},
  \citenamefont {Kurchan}, \citenamefont {Maggs},\ and\ \citenamefont
  {Pujos}}]{gap_first-order3}%
  \BibitemOpen
  \bibfield  {author} {\bibinfo {author} {\bibfnamefont {T.}~\bibnamefont
  {J\"org}}, \bibinfo {author} {\bibfnamefont {F.}~\bibnamefont {Krzakala}},
  \bibinfo {author} {\bibfnamefont {J.}~\bibnamefont {Kurchan}}, \bibinfo
  {author} {\bibfnamefont {A.~C.}\ \bibnamefont {Maggs}},\ and\ \bibinfo
  {author} {\bibfnamefont {J.}~\bibnamefont {Pujos}},\ }\bibfield  {title}
  {\bibinfo {title} {Energy gaps in quantum first-order
  mean-field{\textendash}like transitions: The problems that quantum annealing
  cannot solve},\ }\href {https://doi.org/10.1209/0295-5075/89/40004}
  {\bibfield  {journal} {\bibinfo  {journal} {EPL}\ }\textbf {\bibinfo {volume}
  {89}},\ \bibinfo {pages} {40004} (\bibinfo {year}
  {2010}{\natexlab{b}})}\BibitemShut {NoStop}%
\bibitem [{\citenamefont {Gross}\ \emph {et~al.}(1985)\citenamefont {Gross},
  \citenamefont {Kanter},\ and\ \citenamefont {Sompolinsky}}]{Potts_glass}%
  \BibitemOpen
  \bibfield  {author} {\bibinfo {author} {\bibfnamefont {D.~J.}\ \bibnamefont
  {Gross}}, \bibinfo {author} {\bibfnamefont {I.}~\bibnamefont {Kanter}},\ and\
  \bibinfo {author} {\bibfnamefont {H.}~\bibnamefont {Sompolinsky}},\
  }\bibfield  {title} {\bibinfo {title} {Mean-field theory of the potts
  glass},\ }\href {https://doi.org/10.1103/PhysRevLett.55.304} {\bibfield
  {journal} {\bibinfo  {journal} {Phys. Rev. Lett.}\ }\textbf {\bibinfo
  {volume} {55}},\ \bibinfo {pages} {304} (\bibinfo {year} {1985})}\BibitemShut
  {NoStop}%
\bibitem [{\citenamefont {Sherrington}\ and\ \citenamefont
  {Kirkpatrick}(1975)}]{SK_model1}%
  \BibitemOpen
  \bibfield  {author} {\bibinfo {author} {\bibfnamefont {D.}~\bibnamefont
  {Sherrington}}\ and\ \bibinfo {author} {\bibfnamefont {S.}~\bibnamefont
  {Kirkpatrick}},\ }\bibfield  {title} {\bibinfo {title} {Solvable model of a
  spin-glass},\ }\href {https://doi.org/10.1103/PhysRevLett.35.1792} {\bibfield
   {journal} {\bibinfo  {journal} {Phys. Rev. Lett.}\ }\textbf {\bibinfo
  {volume} {35}},\ \bibinfo {pages} {1792} (\bibinfo {year}
  {1975})}\BibitemShut {NoStop}%
\bibitem [{\citenamefont {Sherrington}\ and\ \citenamefont
  {Kirkpatrick}(1978)}]{SK_model2}%
  \BibitemOpen
  \bibfield  {author} {\bibinfo {author} {\bibfnamefont {D.}~\bibnamefont
  {Sherrington}}\ and\ \bibinfo {author} {\bibfnamefont {S.}~\bibnamefont
  {Kirkpatrick}},\ }\bibfield  {title} {\bibinfo {title} {Infinite-ranged
  models of spin-glasses},\ }\href {https://doi.org/10.1103/PhysRevB.17.4384}
  {\bibfield  {journal} {\bibinfo  {journal} {Phys. Rev. B}\ }\textbf {\bibinfo
  {volume} {17}},\ \bibinfo {pages} {4384} (\bibinfo {year}
  {1978})}\BibitemShut {NoStop}%
\bibitem [{\citenamefont {Suzuki}(1976)}]{ST_formula}%
  \BibitemOpen
  \bibfield  {author} {\bibinfo {author} {\bibfnamefont {M.}~\bibnamefont
  {Suzuki}},\ }\bibfield  {title} {\bibinfo {title} {Relationship between
  d-dimensional quantal spin systems and (d+1)-dimensional ising systems:
  Equivalence, critical exponents and systematic approximants of the partition
  function and spin correlations},\ }\href
  {https://doi.org/10.1143/PTP.56.1454} {\bibfield  {journal} {\bibinfo
  {journal} {Prog. Theor. Phys.}\ }\textbf {\bibinfo {volume} {56}},\ \bibinfo
  {pages} {1454} (\bibinfo {year} {1976})}\BibitemShut {NoStop}%
\bibitem [{\citenamefont {Edwards}\ and\ \citenamefont
  {Anderson}(1975)}]{replica_trick}%
  \BibitemOpen
  \bibfield  {author} {\bibinfo {author} {\bibfnamefont {S.~F.}\ \bibnamefont
  {Edwards}}\ and\ \bibinfo {author} {\bibfnamefont {P.~W.}\ \bibnamefont
  {Anderson}},\ }\bibfield  {title} {\bibinfo {title} {Theory of spin
  glasses},\ }\href {https://doi.org/10.1088/0305-4608/5/5/017} {\bibfield
  {journal} {\bibinfo  {journal} {J. Phys. F: Met. Phys.}\ }\textbf {\bibinfo
  {volume} {5}},\ \bibinfo {pages} {965} (\bibinfo {year} {1975})}\BibitemShut
  {NoStop}%
\bibitem [{\citenamefont {Thirumalai}\ \emph {et~al.}(1989)\citenamefont
  {Thirumalai}, \citenamefont {Li},\ and\ \citenamefont
  {Kirkpatrick}}]{SK_replica}%
  \BibitemOpen
  \bibfield  {author} {\bibinfo {author} {\bibfnamefont {D.}~\bibnamefont
  {Thirumalai}}, \bibinfo {author} {\bibfnamefont {Q.}~\bibnamefont {Li}},\
  and\ \bibinfo {author} {\bibfnamefont {T.~R.}\ \bibnamefont {Kirkpatrick}},\
  }\bibfield  {title} {\bibinfo {title} {Infinite-range ising spin glass in a
  transverse field},\ }\href {https://doi.org/10.1088/0305-4470/22/16/023}
  {\bibfield  {journal} {\bibinfo  {journal} {J. Phys. A}\ }\textbf {\bibinfo
  {volume} {22}},\ \bibinfo {pages} {3339} (\bibinfo {year}
  {1989})}\BibitemShut {NoStop}%
\bibitem [{\citenamefont {Hubbard}(1959)}]{Hubbard_transform}%
  \BibitemOpen
  \bibfield  {author} {\bibinfo {author} {\bibfnamefont {J.}~\bibnamefont
  {Hubbard}},\ }\bibfield  {title} {\bibinfo {title} {Calculation of partition
  functions},\ }\href {https://doi.org/10.1103/PhysRevLett.3.77} {\bibfield
  {journal} {\bibinfo  {journal} {Phys. Rev. Lett.}\ }\textbf {\bibinfo
  {volume} {3}},\ \bibinfo {pages} {77} (\bibinfo {year} {1959})}\BibitemShut
  {NoStop}%
\end{thebibliography}%

\end{document}